\documentclass[12pt,draftclsnofoot, onecolumn]{IEEEtran}

\usepackage{times}
\usepackage{amsmath,dsfont}
\usepackage{amssymb,amsthm}
\usepackage{epsfig,verbatim}
\usepackage{subfigure}
\usepackage{setspace}
\usepackage{color}
\usepackage{cite}
\usepackage{epstopdf}
\usepackage{graphics}
\usepackage{accents}
\usepackage{acronym}
\usepackage[bookmarks,colorlinks]{hyperref}
\usepackage{booktabs}
\usepackage{mathtools}
\usepackage{algorithm}
\usepackage{algorithmic}
\usepackage{enumitem}

\theoremstyle{plain}
\newtheorem{theorem}{Theorem}

\newtheorem{lemma}[theorem]{Lemma}

\theoremstyle{definition}
\newtheorem{definition}[theorem]{Definition}

\theoremstyle{definition}

\newcommand{\includefig}[1]{\includegraphics[width = 0.65\columnwidth]{#1} 	\vspace{-0.2cm}}
\newcommand{\myVec}[1]{{\boldsymbol{#1}}}
\newcommand{\myMat}[1]{{\boldsymbol{#1}}}
\newcommand{\mySet}[1]{\mathcal{#1}}

\newcommand{\E}{\mathbb{E}}		 			
\newcommand{\myX}{{\myVec{x}}}			 	
\newcommand{\myZ}{{\myVec{z}}}			 	
\newcommand{\myS}{{\myVec{s}}}			 	
\newcommand{\LmmseMat}{\myMat{\Gamma}}		
\newcommand{\lenX}{n}			 			
\newcommand{\lenZ}{p}			 			
\newcommand{\lenS}{k}			 			
\newcommand{\Quan}[2]{Q_{#1}^{#2}}			
\newcommand{\myI}{{\myMat{I}}}			 		
\newcommand{\Rate}{R}			 		
\newcommand{\TilM}{\tilde{M}}
\newcommand{\DynRange}{\gamma}
\newcommand{\DynInt}{\Delta}
\newcommand{\myA}{\myMat{A}}
\newcommand{\myB}{\myMat{B}}
\newcommand{\CovMat}[1]{\myMat{\Sigma}_{#1}}
\newcommand{\opt}{^{\rm o}}			
\newcommand{\myEta}{\eta}
\newcommand{\MyKappa}{\kappa}
\newcommand{\eig}[1]{\lambda_{#1}}			
\newcommand{\Wlevel}{\zeta}
\newcommand{\NetMap}{\psi_{\myVec{\theta}}}

\newcommand{\Pdf}[1]{f_{ #1}}

\acrodef{adc}[ADC]{analog-to-digital convertor}
\acrodef{cs}[CS]{compressed sensing}
\acrodef{dtft}[DTFT]{discrete-time Fourier transform}
\acrodef{dft}[DFT]{discrete Fourier transform}
\acrodef{dnn}[DNN]{deep neural network} 
\acrodef{csi}[CSI]{channel state information}
\acrodef{map}[MAP]{maximum a-posteriori probability}
\acrodef{snr}[SNR]{signal-to-noise ratio}
\acrodef{bs}[BS]{base station} 
\acrodef{iot}[IOT]{Interent of Things}
\acrodef{mimo}[MIMO]{multiple-input multiple-output}
\acrodef{mse}[MSE]{mean-squared error}
\acrodef{mmse}[MMSE]{minimum \ac{mse}}
\acrodef{pdf}[PDF]{probability density function}
\acrodef{rv}[RV]{random variable}
\acrodef{fec}[FEC]{forward error correction}
\acrodef{rs}[RS]{Reed-Solomon}
\acrodef{lti}[LTI]{linear time-invariant}
\acrodef{wss}[WSS]{wide-sense stationary}
\acrodef{psd}[PSD]{power spectral density}
\acrodef{ser}[SER]{symbol error rate} 
\acrodef{ber}[BER]{bit error rate} 
\acrodef{sgd}[SGD]{stochastic gradient descent} 
\acrodef{isi}[ISI]{intersymbol interference}  
\acrodef{awgn}[AWGN]{additive white Gaussian noise} 
\acrodef{ut}[UT]{user terminal} 
\acrodef{mmw}[mmWave]{millimeter wave}
\acrodef{cfl}[CFL]{clustered \ac{fl}} 
\acrodef{fl}[FL]{federated learning}
\acrodef{pic}[PIC]{principal inertia compoenent}
\acrodef{ml}[ML]{machine learning} 
\acrodef{dma}[DMA]{dynamic metasurface antenna} 
\acrodef{doa}[DOA]{direction of arrival}

\IEEEoverridecommandlockouts 

\title{Task-Based Quantization with Application to MIMO Receivers}

\author{
	\IEEEauthorblockN{Nir Shlezinger and Yonina C. Eldar 
	}  
	\thanks{This work received funding from the Benoziyo Endowment Fund for the Advancement of Science, the	Estate of Olga Klein -- Astrachan, the European Union’s Horizon 2020 research and innovation program under grant No. 646804-ERC-COG-BNYQ, and  the Israel Science Foundation under grant No. 0100101.
	}
	\thanks{
		N. Shlezinger  and Y. C. Eldar are with the Faculty of Math and CS, Weizmann Institute of Science, Rehovot, Israel (e-mail: nirshlezinger1@gmail.com; yonina@weizmann.ac.il). 	
	}

	\vspace{-1.0cm}
}

\begin{document}
		
	\maketitle
	\pagestyle{plain}
	\thispagestyle{plain}
	
\begin{abstract}
	Multiple-input multiple-output (MIMO) systems are required to communicate reliably at high spectral bands using a large number of antennas, while operating under strict power and cost constraints. In order to meet these constraints, future MIMO receivers are expected to operate with low resolution quantizers, namely, utilize a limited number of bits for representing their observed measurements, inherently distorting the digital representation of the acquired signals. The fact that MIMO receivers use their measurements for some task, such as symbol detection and channel estimation, other than recovering the underlying analog signal, indicates that the distortion induced by bit-constrained quantization can be reduced by designing the acquisition scheme in light of the system task, i.e., by  {\em task-based quantization}. In this work we survey the theory and design approaches to task-based quantization, presenting model-aware designs as well as data-driven implementations. Then, we show how one can implement a task-based bit-constrained MIMO receiver, presenting approaches ranging from conventional hybrid receiver architectures to structures exploiting the dynamic nature of metasurface antennas. 
	This survey narrows the gap between  theoretical  task-based quantization and its implementation in practice, providing concrete algorithmic and hardware design principles for realizing task-based MIMO receivers. 
\end{abstract}

\maketitle
\vspace{-0.2cm}
\section{Introduction}
\vspace{-0.2cm}
	Modern wireless communications systems face a growing set of demands and challenges. Cellular \acp{bs} are required to reliably provide high throughput to an increasing number of \acp{ut}, while maintaining feasible cost and power consumption. An emerging technology to meet these demands is to equip the wireless \acp{bs} with a large number of antenna elements, realizing {\em massive \ac{mimo} communications}. Theoretical studies indicate that substantial  gains in spectral efficiency can be achieved by letting the number of \ac{bs} antennas grow arbitrarily large \cite{marzetta2010noncooperative, shlezinger2018spectral}.  
An additional method to increase the network throughput is to explore the \ac{mmw} frequency range \cite{andrews2014will}, thus overcoming the spectral congestion of traditional wireless bands.  Such \ac{mmw} communications is particularly suitable for massive \ac{mimo} systems: The short wavelengths of \ac{mmw} signals allows packing a large number of antenna elements at a small physical size, and the massive number of elements facilitates directed beamforming which is essential at \ac{mmw} bands.

While the theoretical gains of massive \ac{mimo} systems, particularly when combined with \ac{mmw} transmission, are clear, implementing such systems in practice under strict cost and power constraints is a challenging task. 
A major source of this increased cost are the \ac{adc} components, which allow the analog signals observed by each antenna element to be processed in digital. The power consumption of an \ac{adc} is directly related to the signal bandwidth and the number of bits used for digital representation \cite{walden1999analog, eldar2015sampling}. Consequently, in massive \ac{mimo} systems, where the number of antennas and \acp{adc} operating at high frequency bands is large, limiting the number of bits, thus operating under quantization constraints, is crucial to keep  cost and power consumption feasible~\cite{andrews2014will}.  

Focusing on uplink communications, i.e., when the \ac{bs} acts as the receiver, quantization constraints imply that the \ac{bs} cannot process the channel output directly but rather only an inaccurate distorted digital representation of it. The distortion induced by continuous-to-discrete quantization mappings  degrades the ability to extract information, such as the underlying channel coefficients or the transmitted signal, from the observed channel output. Consequently, methods for channel estimation and symbol detection from quantized outputs are the focus of a large body of work, including, e.g., \cite{mo2017channel,li2017channel,jacobsson2017throughput,choi2016near,pirzadeh2018spectral,khobahi2019deep}. These schemes are carried out in the digital domain, i.e., they are {\em digital-only methods}, assuming a fixed quantization system.

An alternative emerging approach to processing only in the digital domain, which is the focus of the current survey, is to jointly design the quantization system along with the digital processing in light of the task as proposed in \cite{shlezinger2018hardware}. Such {\em task-based quantization} systems convert their received analog signal into a digital representation in a manner which preserves the semantic information required to carry out the task, rather than recovering the analog signal, thus allowing to operate efficiently with standard \acp{adc} under relatively tight bit constraints \cite{shlezinger2018hardware,shlezinger2018asymptotic,salamatian2019task,shlezinger2019deep}. Task-based quantizers,  originally derived for generic digital signal processing applications in \cite{shlezinger2018hardware}, bear the potential of significantly facilitating the design of massive \ac{mimo} receivers operating under bit constraints \cite{shlezinger2018asymptotic}. This follows since in \ac{mimo} systems, acquisition is carried out for specific tasks, most commonly channel estimation  and symbol detection. These can be treated as recovering information embedded in the received signals, which in turn can be accurately and compactly extracted in digital using task-based quantization.

In this work we survey recent results in task-based quantization. We focus on its application for bit-constrained \ac{mimo} receivers, although task-based quantization is relevant in many other applications including sensor arrays, radar, medical imaging, and essentially any system which acquires physical signals for some task while operating under bit constraints. We begin by detailing model-aware methods for designing task-based quantizers. These methods jointly design the overall acquisition system along with the digital processing based on prior knowledge of the statistical model relating the observed analog signal and the desired task information to be extracted in digital. Our model-aware analysis characterizes the achievable accuracy in recovering the desired information under bit constraints for tasks which can be modeled as a linear function of the measurements as in, e.g.,  Rayleigh fading \ac{mimo}  channel estimation \cite{shlezinger2018asymptotic}. Then, we show how the proposed approach can be extended to more involved tasks by utilizing the mathematical tool of \acp{pic} \cite{du2017principal}. Specifically, we show that \acp{pic} can facilitate identifying a proper transformation of the measurements from which the task can be treated as approximately linear, allowing to use the proposed task-based quantizer. We specialize the derivation for tasks where the desired information is encapsulated in quadratic functions of the measurements, which is the case in, e.g., covariance estimation \cite{rodrigues2017rate} and \ac{doa} recovery \cite{yu2016doa}.   

Next, we show how  task-based quantization systems can be designed without explicitly specifying the statistical relationship between the observations and the desired task information, by tuning the overall acquisition system in a data-driven manner. We demonstrate how by combining \ac{ml} tools with an accurate differentiable approximation of the quantization rule, one can learn task-based quantization mappings from a set of labeled data. Finally, we show how to implement \ac{mimo} receivers capable of dynamically adjusting their acquisition system in light of the task, thus realizing tunable task-based quantization. Our proposed design builds upon either conventional hybrid receiver architectures \cite{mendez2016hybrid,ioushua2019family}, dedicated pre-acquisition hardware \cite{gong2019rf}, or on exploiting the inherent configurability of receivers equipped with metasurface antennas \cite{shlezinger2019dynamic,wang2019dynamic}, and we present hardware prototypes built in our lab, demonstrating the feasibility of task-based quantization in \ac{mimo} receivers.


The rest of this paper is organized as follows: 
Section~\ref{sec:Rx} formulates the system model and reviews some basics in quantization theory.
Methods for designing task-based quantizers based on prior model knowledge are detailed in Section~\ref{sec:Task}. Section~\ref{sec:TaskDeep} presents data-driven design strategies.  
In Section~\ref{sec:Practice} we show how one can implement  task-based quantization in bit-constrained \ac{mimo} receivers, reviewing several candidate architectures and hardware prototypes.   
Section~\ref{sec:Conclusion} provides some concluding remarks.

Throughout the paper, we use boldface lower-case letters for vectors, e.g., ${\myVec{x}}$,
where the $i$th element of ${\myVec{x}}$ is written as $({\myVec{x}})_i$.
Boldface upper-case letters are used for matrices,  e.g., $\myMat{M}$,  
where  $(\myMat{M})_{i,j}$   denotes its $(i,j)$th element. 
Sets are denoted with calligraphic letters, e.g., $\mathcal{X}$. 
We use $\myI_{n}$ to represent the $n \times n$ identity matrix. 
Transpose,  Euclidean norm, Kronecker product, and stochastic expectation are written as  $(\cdot)^T$,  $\left\|\cdot\right\|$, $\otimes$, and $\E\{ \cdot \}$,   respectively, and
$\mySet{R}$ is the set of real numbers. 
All logarithms are taken to basis two.

\vspace{-0.2cm}
\section{Preliminaries and Problem Formulation}
\label{sec:Rx}

\vspace{-0.2cm}
\subsection{Preliminaries in Quantization Theory}
\label{subsec:RxQuant}
\begin{figure}
	\centering
	\includefig{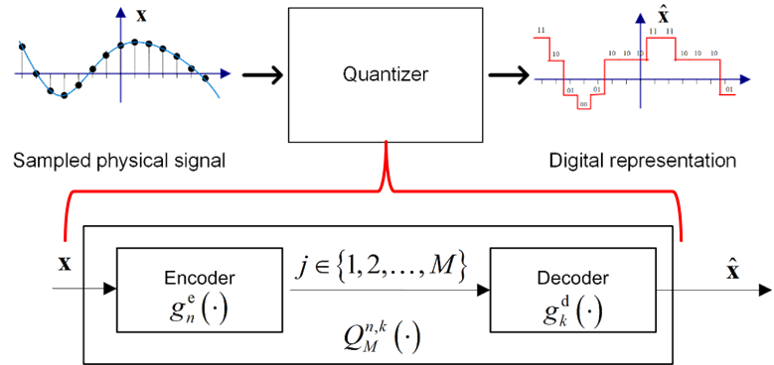} 
	\caption{Quantizer illustration.} 
	\label{fig:Quantizer1}
\end{figure}

We begin by briefly reviewing the standard quantization setup, and recall the definition of a quantizer:
\begin{definition}[Quantizer]
	\label{def:Quantizer}
	A quantizer $\Quan{M}{n,k}\left(\cdot \right)$ with $\log M$ bits, input size $n$, input alphabet $\mySet{X}$, output size $k$, and output alphabet $\hat{\mySet{X}}$, consists of: 
	{\em 1)} An  encoding function $g_n^{\rm e}: \mySet{X}^n \mapsto \{1,2,\ldots,M\} \triangleq \mySet{M}$ which maps the input into a discrete index.
	{\em 2)} A decoding function  $g_k^{\rm d}: \mySet{M} \mapsto \hat{\mySet{X}}^k$ which maps each index $j \in \mySet{M}$ into a codeword $\myVec{q}_j \in  \hat{\mySet{X}}^k$. 
\end{definition}
\noindent
We write the output of the quantizer with input $\myX  \in \mySet{X}^n$ as $\hat{\myX}  = g_k^{\rm d}\left( g_n^{\rm e}\left( \myX \right) \right) \triangleq \Quan{M}{n,k}\left( \myX \right)$. 
{\em Scalar quantizers} operate on a scalar input, i.e., $n=1$ and $\mySet{X}$ is a scalar space, while {\em vector quantizers} have a multivariate input.  An illustration of a quantization system is depicted in Fig. \ref{fig:Quantizer1}.

In the standard quantization problem, a $\Quan{M}{n,n}\left(\cdot \right)$ quantizer is designed to minimize some distortion measure  $d:\mySet{X}^n\times\hat{\mySet{X}}^n \mapsto \mySet{R}^+$  between its input and its output. 
The performance of a quantizer is  characterized using its quantization rate $\Rate \triangleq \frac{1}{n}\log M$, and the expected distortion $\E\{d\left(\myX , \hat{\myX}  \right)\}$. For a fixed input size $n$ and codebook size $M$, the optimal quantizer is
$\Quan{M}{n, {\rm opt}}\left(\cdot \right) = \mathop{\arg \min}_{\Quan{M}{n,n}} \E \left\{d\left(\myX, \Quan{M}{n,n}\left( {\myX} \right)\right)   \right\}$.
Characterizing the optimal quantizer and its trade-off between distortion and quantization rate is in general a very difficult task. Optimal quantizers are thus typically studied assuming either high quantization rate, i.e., $\Rate \rightarrow \infty$, see, e.g., \cite{gray1998quantization}, or asymptotically large inputs, namely, $n \rightarrow \infty$, via rate-distortion theory \cite[Ch. 10]{cover2012elements}. 


\vspace{-0.2cm}
\subsection{Problem Formulation}
\label{subsec:Problem}
\vspace{-0.2cm}
Here, we study {\em task-based quantization} \cite{shlezinger2018hardware}, where the design objective of the quantizer is some task 
other than minimizing the distortion between its input and output. 
In the following, we focus on the generic task of acquiring a random vector  $\myS \in \mySet{S}^\lenS \subseteq \mySet{R}^\lenS$ from a  statistically dependent random vector $\myX \in \mySet{R}^\lenX$ of larger dimensionality, i.e., $\lenX \geq \lenS$.  The set $\mySet{S}$ represents the possible values of the unknown vector: It can be continuous, representing an estimation task; or discrete, for classification tasks.
This formulation accommodates a broad range of applications, including channel estimation and symbol detection, that are the common tasks considered in  \ac{mimo} communications receivers \cite{shlezinger2018asymptotic}, as well as covariance recovery \cite{rodrigues2017rate}, \ac{doa} estimation \cite{yu2016doa}, and source localization \cite{corey2017wideband}.  
 The recovered estimate of $\myS$, denoted $\hat{\myS}$, is represented in digital using up to $\log M$ bits, dictating the bit budget allowed for task-based quantization. The observed $\myX$ is related to $\myS$ via a conditional probability measure $\Pdf{\myX|\myS}$. For example, in a communications setup. the conditional probability measure $\Pdf{\myX|\myS}$ encapsulates the noisy channel. 

The performance limits of task-based quantization with asymptotically large vectors, i.e., when $\lenX \rightarrow \infty$ while $\Rate = \frac{1}{n}\log M$ remains fixed, can be characterized using indirect rate-distortion theory \cite{witsenhausen1980indirect}. Specifically, for estimation tasks with the \ac{mse} distortion objective, i.e., $d(\myVec{s}, \hat{\myVec{s}}) = \| \myVec{s} - \hat{\myVec{s}} \|^2 $, the task-based quantization mapping which minimizes the \ac{mse} for a fixed quantization rate $\Rate$ was derived in \cite{wolf1970transmission} for fixed-size vectors. The resulting optimal strategy consists of applying vector quantization to the \ac{mmse} estimate of  $\myVec{s}$ from  $\myVec{x}$. 

While vector quantizers allow to achieve more accurate digital representations of the acquired analog signal compared to their scalar counterparts \cite[Ch. 23]{polyanskiy2015lecture}, practical \acp{adc} typically utilize scalar quantizers. In particular, \acp{adc} often apply the same continuous-to-discrete mapping to each sample, which is most commonly based on a uniform partition of the real line, i.e., scalar uniform quantization \cite{walden1999analog}. Nonetheless, in the presence of a task, one is not interested in recovering the analog signal, but rather estimate some underlying information embedded in it. This motivates the analysis of how to incorporate the presence of a task in the design of a quantization system utilizing scalar \acp{adc}, and whether the distortion induced by conventional scalar quantization can be mitigated when recovering the task.

%
\begin{figure}
	\centering
	\includegraphics[width=14cm]{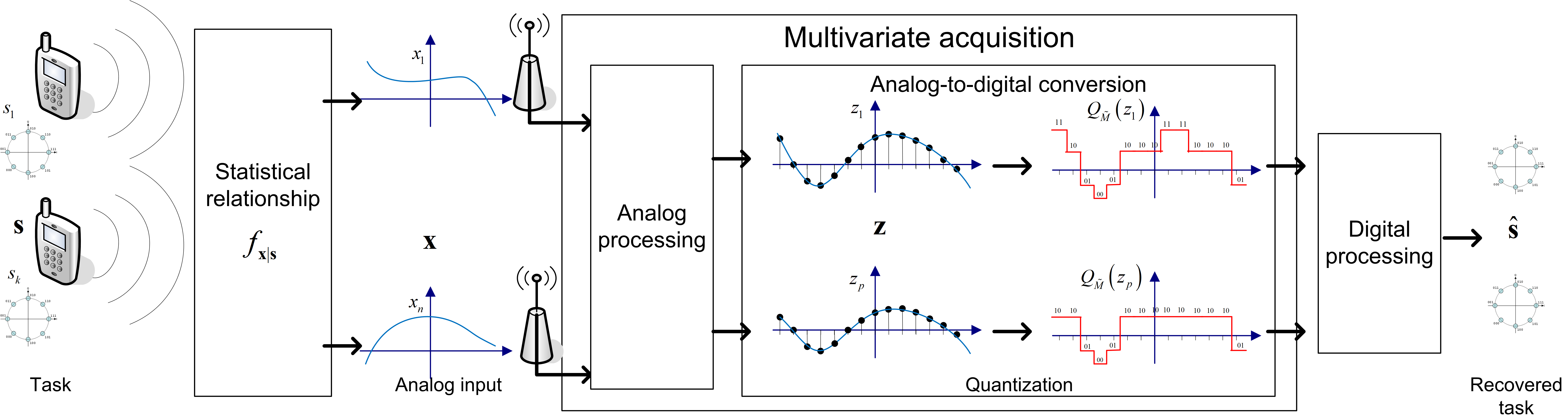} 
	\caption{Hybrid quantization system model. For illustration, the task is recovering a set of constellation symbols in uplink MIMO communications.} 
	\label{fig:SystemModel}
\end{figure}
%

\vspace{-0.2cm}
\subsection{Hardware-Limited Task-Based Quantization}
\label{subsec:RxHybrid}
\vspace{-0.2cm}
As discussed in the previous section, practical digital signal processing systems typically obtain a discrete representation of physical analog signals using  scalar \acp{adc}. 
In such systems, each continuous-amplitude sample is converted into a discrete representation using a single quantization rule. Therefore, in order to be able to account for the presence of a task in acquisition while operating with scalar \acp{adc}, one must introduce some level of processing, in addition to that carried out in digital. We therefore consider  hybrid acquisition systems as illustrated in Fig. \ref{fig:SystemModel}, which is a common model in \ac{mimo} communication receivers \cite{mendez2016hybrid,ioushua2019family}. Hybrid architectures were originally proposed as a method to reduce the number of costly RF chains in \ac{mimo} receivers \cite{mendez2016hybrid,ioushua2019family}, while here we exploit these structures to allow quantization under bit constraints for tasks. 
In such hybrid systems, a set of analog signals can be combined in analog prior to being converted to digital, a property which we exploit in order to facilitate extracting some desired information from them. This model can represent, e.g., sensor arrays or \ac{mimo} receivers, and specializes the case of a single analog input signal. While acquiring a set of analog signals in digital hardware includes both sampling, i.e., continuous-to-discrete time conversion, as well as quantization, we henceforth focus only the quantization aspect assuming a fixed sampling mechanism. The joint design of sampling and quantization in light of a task is left for future studies; initial results can be found in \cite{shlezinger2020learning}.

\begin{figure}
	\centering
	\includegraphics[width=12cm]{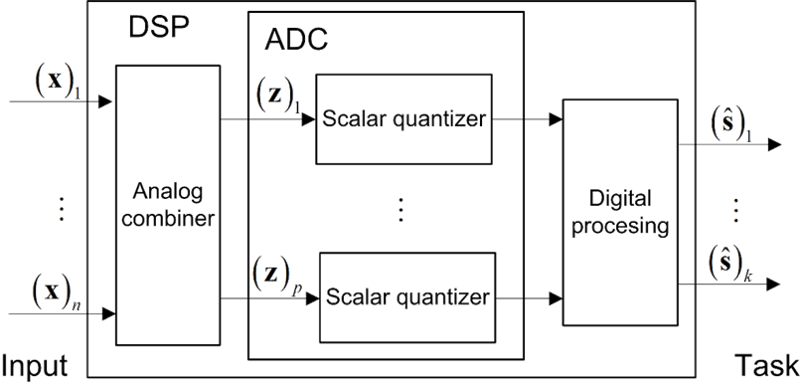} 
	\vspace{-0.2cm}
	\caption{Block diagram of considered task-based quantization systems.} 
	\label{fig:HybridSystem2}
\end{figure}

In the proposed hybrid architecture, the input to the \ac{adc}, denoted $\myZ \in \mySet{R}^\lenZ$, where $\lenZ$ denotes the number of scalar quantizers, is obtained from $\myX$ using a pre-quantization mapping  referred to as {\em analog combining}. Then, $\myZ$ is quantized using  $\lenZ$ identical scalar quantizers with resolution $\TilM \triangleq \lfloor M^{1/\lenZ}\rfloor$ into a digital vector $Q(\myZ)$. The overall number of bits is $\lenZ \cdot \log \TilM \le \log M$. The \ac{adc} output is processed in  digital  to obtain the estimate $\hat{\myS} \in \mySet{S}^\lenS$. A schematic block diagram of the quantization system is depicted in Fig. \ref{fig:HybridSystem2}. Designing task-based quantizers can be formulated as the joint optimization of the analog combining mapping, the scalar quantization rule, and the digital processing, such that the output $\hat{\myS}$ will be an accurate estimate of the task vector $\myS$, while operating under a fixed budget of up to $\log M$ bits. 

The characterization of task-based quantization systems of the form of Fig. \ref{fig:HybridSystem2} consists of two complementary studies: First, we study in Section \ref{sec:Task} how the overall system can be designed based on knowledge of the conditional distribution $\Pdf{\myX|\myS}$ relating the observations and the task in a model-based fashion. Then, we discuss how task-based quantization mappings can be learned from labeled data building upon \ac{ml} tools, and in particular, by utilizing \acp{dnn} to adapt task-based quantization mappings,  in Section \ref{sec:TaskDeep}. Our results demonstrate that by properly tuning the hybrid architecture of task-based quantizers, one can approach the performance limits dictated by indirect rate-distortion theory, achievable using complex vector quantizers, while using conventional scalar \acp{adc} operating as part of an acquisition system of feasible hardware requirements.

\vspace{-0.2cm}
\section{Model-Aware Task-Based Quantization}
\label{sec:Task}
\vspace{-0.2cm}
In this section we detail how to design hybrid quantization systems to facilitate the recovery of the task vector $\myS$ in the digital domain, based on prior knowledge of the underlying statistical model. In particular, we discuss how the analog combining, quantization rule, and digital processing components of the system in Fig. \ref{fig:HybridSystem2} can be jointly optimized based on knowledge of the conditional distribution relating the input $\myX$ to the task vector $\myS$, denoted $\Pdf{\myX | \myS}$. We begin by presenting the model assumptions under which the analysis is carried out in Section \ref{subsec:TaskModel}. After that we present the resulting task-based quantization systems for estimation tasks of linear and quadratic nature in Sections \ref{subsec:TaskLinear}-\ref{subsec:TaskQuadratic}, respectively. 

\vspace{-0.2cm}
\subsection{System Model}
\label{subsec:TaskModel}
\vspace{-0.2cm}
	In order to obtain a meaningful and tractable characterization of the task-based quantization  system of Fig. \ref{fig:HybridSystem2}, we henceforth introduce two model assumptions upon which we base our results in the remainder of this section: 
		\begin{enumerate}[label={\em A\arabic*}]
			\item \label{itm:A1} We consider the task of estimating the task $\myS$ in the \ac{mse} sense, namely, our performance measure is the \ac{mse} $\E\{\|\myS - \hat{\myS}\|^2 \} $.
			\item \label{itm:A2} We focus on uniform \acp{adc}, and model the their operation in our derivations as non-subtractive uniform dithered quantizers \cite{gray1993dithered}. 
		\end{enumerate}
	
	Model assumption \ref{itm:A1} implies that the fidelity of an estimate $\hat{\myS}$ can be represented as a sum of the \ac{mmse} and the excess \ac{mse} with  respect to the \ac{mmse} estimate $\tilde{\myS} = \E\{\myS|\myX\}$, as $\E\{\|\myS - \hat{\myS}\|^2 \} = \E\{\|\myS - \tilde{\myS} \|^2 \} + \E\{\|\tilde{\myS}  - \hat{\myS}\|^2 \}$. Consequently, in the following we characterize the performance in terms of the excess \ac{mse} $ \E\{\|\tilde{\myS}  - \hat{\myS}\|^2 \}$. Since $\tilde{\myS}$ is a function of $\myX$, we divide our analysis based on the nature of this function, considering linear functions in Section \ref{subsec:TaskLinear}, extending to quadratic and more general forms in Section~\ref{subsec:TaskQuadratic}.  
	
Model assumption \ref{itm:A2} imposes a structure on the scalar quantization mapping. 
To formulate the resulting input-output relationship of the \acp{adc}, let $\DynRange$ denote the support of the quantizer, and define $\DynInt \triangleq \frac{2\DynRange}{\TilM}$ as the quantization spacing. 
The output of the uniform \ac{adc} with input sequence $z_1, z_2, \ldots, z_\lenZ$ can be written as $Q\left( z_i\right)  =  q\left(z_i + u_i \right) $, where $u_1, u_2, \ldots, u_\lenZ$ are i.i.d. \acp{rv} uniformly distributed over $\left[-\frac{\DynInt}{2},\frac{\DynInt}{2} \right]$, mutually independent of the input, representing the dither signal.
The function $q(\cdot)$, which implements the uniform quantization, is given by  
\vspace{-0.2cm}
\begin{equation}
q(z) = \begin{cases}
-\DynRange + \DynInt\left(l - \frac{1}{2} \right)   & 
z - l  \DynInt \in  \left[-\frac{\DynInt}{2},\frac{\DynInt}{2} \right],  l \in \{0,1, \ldots, \TilM - 1 \}  \\
{\rm sign}\left(z\right) \left( \DynRange - \frac{\DynInt}{2}\right)    & |z| > \DynRange.
\end{cases}
\label{eqn:UniQuant}
\vspace{-0.2cm}
\end{equation}   
When $\TilM = 2$, the resulting quantizer is a standard one-bit sign quantizer of the form $q(z) = c \cdot {\rm sign}(z)$, where $c >0$ is determined by the support $\DynRange$.

	Dithered quantizers significantly facilitate the analysis, due to the  following favorable property: When operating within the support, the output can be written as the sum of the input and an additive zero-mean white quantization noise signal uncorrelated with the input. The drawback of adding dither is that it increases the energy of the quantization noise, namely, it results in increased distortion \cite{gray1993dithered}. Nonetheless, the favorable property of dithered quantization is also satisfied in uniform quantization  {\em without dithering} for inputs with bandlimited characteristic functions, and is approximately satisfied for various families of input distributions \cite{widrow1996statistical}. Consequently, while our  analysis assumes dithered quantization, exploiting  the resulting statistical properties of the quantization noise, the proposed system is applicable without dithering, as we demonstrate in our numerical study. 

\vspace{-0.2cm}
\subsection{Linear Estimation Tasks}
\label{subsec:TaskLinear}
\begin{figure}
	\centering
	\includegraphics[width=12cm]{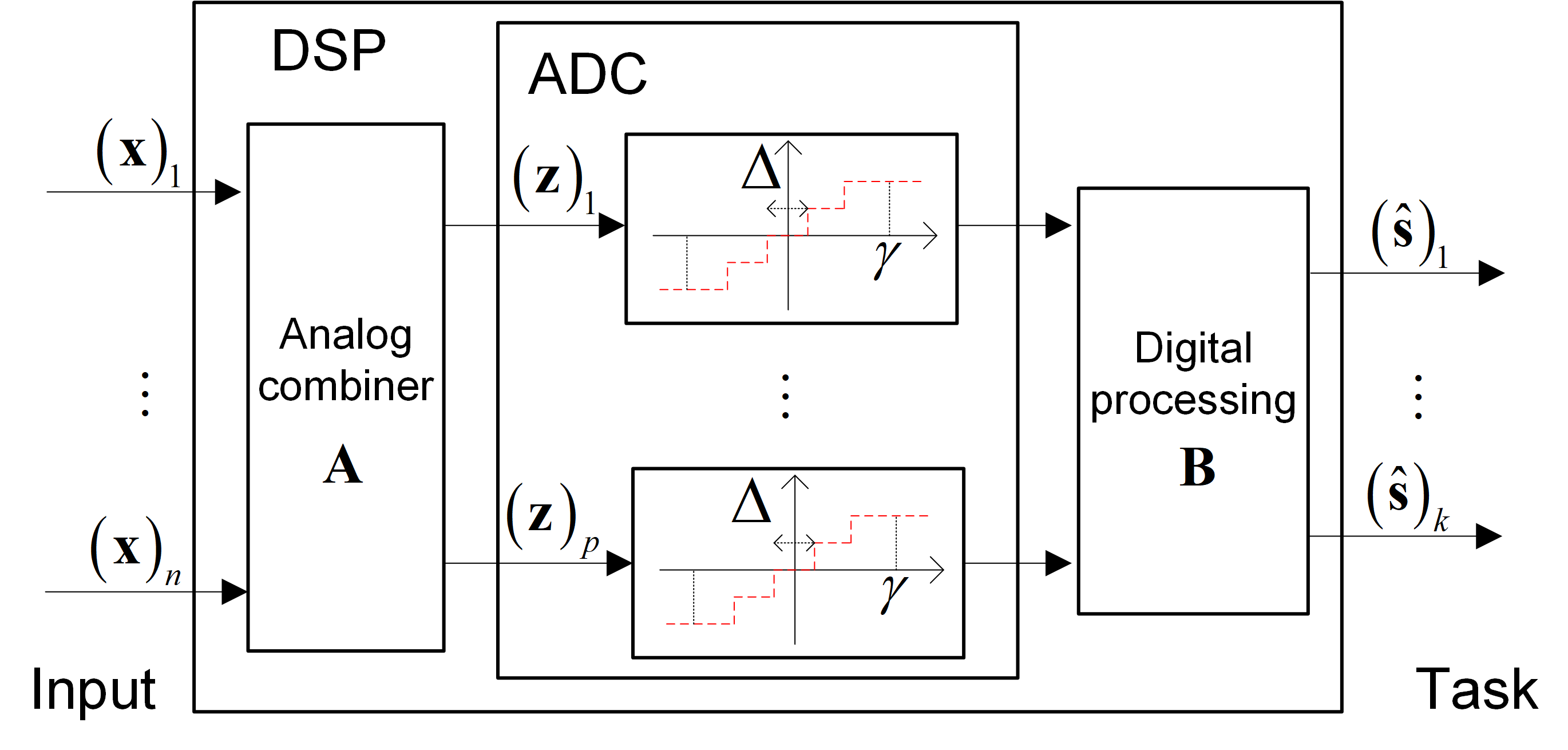} 
	\vspace{-0.2cm}
	\caption{Model-aware task-based quantization for linear tasks illustration.} 
	\label{fig:OverallStructure3}
\end{figure}

\vspace{-0.2cm}
We begin by focusing on scenarios in which the stochastic relationship between the vector of interest $\myS$ and the observations $\myX$ are such that the \ac{mmse} estimate of $\myS$ from $\myX$  is a linear function of $\myX$, i.e., $\exists \LmmseMat \in \mySet{R}^{\lenS \times \lenX}$ such that $\tilde{\myS} = \LmmseMat \myX$.  	
Accordingly,  we restrict the analog combining and the digital mapping components in Fig. \ref{fig:HybridSystem2} to be linear, namely, $\myZ = \myA \myX$ and $\hat{\myS} = \myB Q(\myZ)$, for some $\myA \in \mySet{R}^{\lenZ \times \lenX}$ and  $\myB \in \mySet{R}^{\lenS \times \lenZ}$. 
An illustration of the considered system architecture is depicted in Fig. \ref{fig:OverallStructure3}. 
%
%
%
By focusing on these setups, we are able to explicitly derive the achievable distortion and to characterize the system which minimizes the \ac{mse}. This derivation reveals some non-trivial insights. For example, we show that the optimal approach when using vector quantizers, namely, to quantize the \ac{mmse} estimate \cite{wolf1970transmission}, is no longer optimal when using  standard scalar \acp{adc}. Furthermore, as detailed in Section \ref{subsec:TaskQuadratic}, our analysis provides  guidelines for designing  task-based quantization systems which can be used for more general relationships between $\myS$ and $\myX$, such as the recovery of quadratic tasks.

Let $\CovMat{\myX}$ be the covariance matrix of $\myX$, assumed to be non-singular.
Before we study the overall task-based quantization system, we first derive the  digital processing matrix which minimizes the \ac{mse} for a given analog combiner $\myA$ and the resulting  \ac{mse},  stated in the following lemma \cite[Lem. 1]{shlezinger2018hardware}:
\begin{lemma}
	\label{lem:ThmProof1}
	\begin{subequations}
	For any analog combining matrix $\myA$ and support $\DynRange$ such that the quantizers operate within their support, i.e., $\Pr \left( \big|\left( \myA \myX\right)_l + u_l\big| > \DynRange \right) = 0$,   
	the  digital processing matrix which minimizes the \ac{mse} is given by 
	\begin{equation}
	\myB\opt \left( \myA\right)  = \LmmseMat\CovMat{\myX}\myA^T\bigg( \myA\CovMat{\myX}\myA^T + \frac{{2{\DynRange^2}}}{{3\TilM^2}}{\myI_\lenZ} \bigg)^{ - 1},
	\end{equation}
	and the achievable excess \ac{mse}, denoted ${\rm MSE}\left(\myA\right) 
	= \mathop {\min }_{\myB} \E\big\{ \big\| \tilde{\myS}-\hat{\myS} \big\|^2 \big\} $, is 
	\begin{align}
	{\rm MSE}\left(\myA\right) 
	&\!=\! 
	{\rm Tr} \bigg( \LmmseMat \CovMat{\myX}\LmmseMat^T\! -\! \LmmseMat\CovMat{\myX}\myA^T\!\bigg( \myA\CovMat{\myX}\myA^T\! + \!\frac{{2{\DynRange^2}}}{{3\TilM^2}}{{\bf{I}}_\lenZ} \bigg)^{ - 1}\!\!\!\!\!\myA\CovMat{\myX}\LmmseMat^T \bigg).
	\end{align} 
	\end{subequations}	
\end{lemma}

The digital processing matrix in Lemma~\ref{lem:ThmProof1} is the linear  \ac{mmse} estimator of $\myS$ from the vector $\myA\myX + \myVec{e}$, where $\myVec{e}$ represents the quantization noise, which is white and uncorrelated with $\myA\myX$. This stochastic representation is a result of the usage of non-overloaded dithered quantizers. 
Nonetheless, in the following we use the model on which Lemma~\ref{lem:ThmProof1} is based  to design task-based quantizers operating with small yet non-zero probability of overloading, i.e.,  $\Pr \left( \big|\left( \myA \myX\right)_l + u_l\big| > \DynRange \right) \approx 0$ for each $l$. In such cases modeling  $\myA\myX$ and $\myVec{e}$ as uncorrelated becomes a reliable approximation. Therefore, in order to use Lemma \ref{lem:ThmProof1} to design task-based quantizers,  we explicitly require to avoid overloading with high probability. This is achieved by fixing $\DynRange$ to be some multiple $\myEta$ of the maximal standard deviation of the input, allowing to bound the overload probability via  Chebyshev's inequality \cite[Pg. 64]{cover2012elements}. 

We now use Lemma \ref{lem:ThmProof1} to obtain the analog combining matrix $\myA\opt$ which minimizes the \ac{mse} and the resulting system.
Define the matrix $\tilde{\LmmseMat} \triangleq \LmmseMat \CovMat{\myX}^{1/2}$,  let $\{ \eig{\tilde{\LmmseMat},i} \}$ be its singular values arranged in a descending order, and set  $\MyKappa \triangleq \myEta^2 \big(1 - \frac{ \myEta^2 }{3\TilM^2}\big)^{-1}$. Note that for $i > {\rm rank} \big( \tilde{\LmmseMat}\big)$, $\eig{\tilde{\LmmseMat},i}  = 0$. The resulting task-based quantization system is stated in the following theorem \cite[Thm. 1]{shlezinger2018hardware}:
\begin{theorem}
	\label{thm:OptimalDes}
	\begin{subequations}
		\label{eqn:OptimalDes}
		For the task-based quantization system under linear estimation tasks, the analog combining matrix $\myA\opt$ is given by $\myA\opt = \myMat{U}_{\myA} \myMat{\Lambda}_{\myA} \myMat{V}_{\myA}^T \CovMat{\myX}^{-1/2}$, where  $\myMat{V}_{\myA} \in \mySet{R}^{\lenX \times \lenX}$ is the right singular vectors matrix of  $\tilde{\LmmseMat}$; 
		  $\myMat{\Lambda}_{\myA} \in \mySet{R}^{\lenZ \times \lenX}$ is a diagonal matrix with diagonal entries  
			\begin{equation}
			\label{eqn:OptimalDesA}
			\left( \myMat{\Lambda}_{\myA}\right)_{i,i}^2 = \frac{{2{\MyKappa }}}{{3\TilM^2} \cdot \lenZ}\left( {\Wlevel  \cdot\eig{\tilde{\LmmseMat},i} - 1} \right)^ +,
			\end{equation}
			with  $\Wlevel$  set such that $\frac{{2{\MyKappa }}}{{3\TilM^2} \cdot \lenZ}\sum_{i=1}^{\lenZ} \big( {\Wlevel  \cdot\eig{\tilde{\LmmseMat},i} - 1} \big)^ + = 1$; 
			and $\myMat{U}_{\myA} \in \mySet{R}^{\lenZ \times \lenZ}$ is a unitary matrix which guarantees that  
			 $\myMat{U}_{\myA}\myMat{\Lambda}_{\myA}\myMat{\Lambda}_{\myA}^T\myMat{U}_{\myA}^T$ is weakly majorized by all possible rotations of  $\myMat{\Lambda}_{\myA}\myMat{\Lambda}_{\myA}^T$. 
		The support of the \ac{adc} is given by $	\DynRange^2   =   \frac{ \MyKappa}{\lenZ}$, 
		and the digital processing matrix is equal to	
		\begin{equation}
		\label{eqn:OptimalDesB}
		\myB\opt\left( \myA\opt\right) = \tilde{\LmmseMat} \myMat{V}_{\myA}\myMat{\Lambda}_{\myA}^T\left( \myMat{\Lambda}_{\myA} \myMat{\Lambda}_{\myA}^T+ \frac{{2{\DynRange^2}}}{{3\TilM^2}}\myI_{\lenZ} \right)^{ - 1}\!\! \myMat{U}_{\myA}^T.
		\end{equation}
		The resulting minimal achievable excess \ac{mse} is
		\begin{equation}
		\label{eqn:OptimalDesMSE}	
		\E \left\{\left\|\tilde{\myS} \!-\! \hat{\myS} \right\|^2  \right\}\! = \!
		\begin{cases} 
		\sum\limits_{i=1}^{\lenS}   \frac{ \eig{\tilde{\LmmseMat},i}^2} {\left(\Wlevel \cdot\eig{\tilde{\LmmseMat},i} -  1 \right)^+ \!+ \!1}, &\lenZ \!\ge\! \lenS \\
		\sum\limits_{i=1}^{\lenZ}   \frac{ \eig{\tilde{\LmmseMat},i}^2} {\left(\Wlevel \cdot\eig{\tilde{\LmmseMat},i} -  1 \right)^+ \!+\! 1} \!+\! \sum\limits_{i\!=\!\lenZ\!+\!1}^{\lenS}  \eig{\tilde{\LmmseMat},i}^2, & \lenZ\! <\! \lenS.
		\end{cases} 
		\end{equation}
	\end{subequations}
\end{theorem}  

\begin{figure}
	\centering
	\includegraphics[width=14cm]{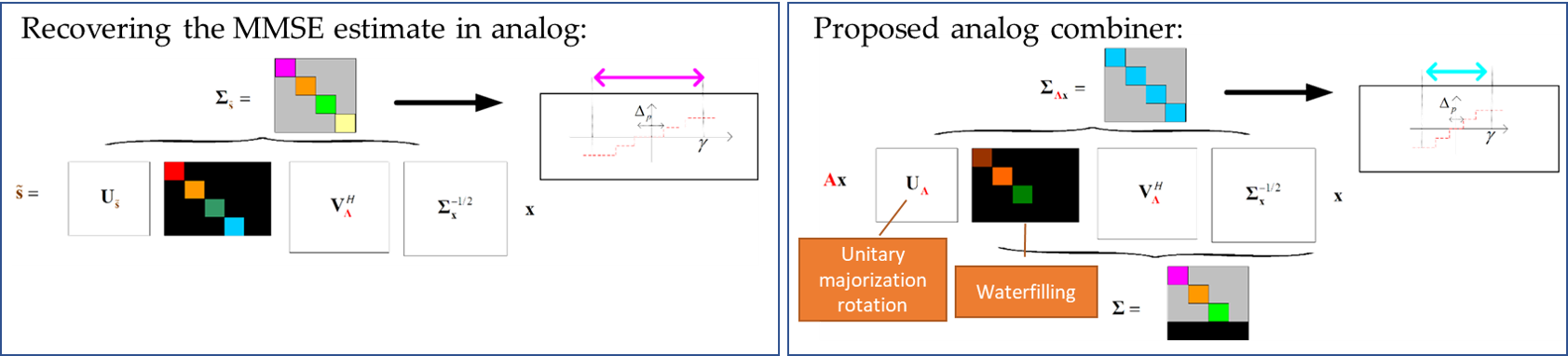} 
	\caption{An illustration of the ADC input, its covariance, and the resulting quantization mapping when quantizing the \ac{mmse} estimate (left) and for the proposed combiner of Theorem \ref{thm:OptimalDes} (right).} 
	\label{fig:ThmIllust}
\end{figure}

The majorizing unitary matrix $\myMat{U}_{\myA}$ is guaranteed to exist by  \cite[Cor. 2.1]{palomar2007mimo}, and can obtained via, e.g., \cite[Alg. 2.2]{palomar2007mimo}. 
Since the design objective is the \ac{mse} by \ref{itm:A1}, the optimal quantization system utilizing vector quantizers is known to recover $\tilde{\myS} = \LmmseMat\myX$ in the analog domain \cite{wolf1970transmission}. In the presence  of scalar \acp{adc}, Theorem \ref{thm:OptimalDes} reveals two main differences in the desired pre-quantization mapping: First, the analog combiner essentially nullifies the weak eigenmodes of the correlation matrix of the \ac{mmse} estimate in \eqref{eqn:OptimalDesA}, as these eigenmodes are likely to become indistinguishable by finite resolution uniform scalar quantization. 
Then, the unitary rotation matrix  $\myMat{U}_{\myA}$, which guarantees that the entries of $\myZ$ have the same variance, minimizes the maximal variance of the quantized variables, allowing to use relatively fine quantization at a given resolution without risking high overloading probability. This combined operation of the analog mapping trades estimation error and quantization accuracy, allowing to optimize the digital representation in light of the task. An illustration of this analog combiner and its quantization rule compared to recovering $\tilde{\myS}$ in analog is depicted in Fig. \ref{fig:ThmIllust}.

The characterization of the task-based quantization system in Theorem~\ref{thm:OptimalDes} gives rise to the following non-trivial insights: 
$1)$ 	In order to minimize the \ac{mse}, $\lenZ$ must not be larger than the rank of the covariance matrix of  $\tilde{\myS}$ \cite[Cor. 1]{shlezinger2018hardware}. This implies that reducing the dimensionality of the input prior to quantization contributes to recovering the task vector as higer resolution quantizers can be used without violating the overall bit constraint;
and $2)$ 	When the covariance matrix of $\tilde{\myS}$ is non-singular, quantizing the \ac{mmse} estimate minimizes the \ac{mse} if and only if the covariance matrix of  $\tilde{\myS}$ equals   $\myI_{\lenS}$ up to a constant factor \cite[Cor. 4]{shlezinger2018hardware}. 
This indicates that, except for very specific statistical models, quantizing the entries of the \ac{mmse} estimate vector, which is the optimal strategy when using vector quantizers \cite{wolf1970transmission}, does not minimize the \ac{mse} when using uniform scalar \acp{adc}. 

To illustrate the gains of the task-based quantization system design which arises from Theorem \ref{thm:OptimalDes}, we next numerically evaluate its achievable \ac{mse} in a simulation study. 
We  consider the estimation of a scalar \ac{isi} channel from quantized observations.
In this scenario, the parameter vector $\myS$ represents the coefficients of a multipath channel  with $\lenS$ taps.
The channel is estimated from a set of $\lenX = 120$ noisy observations $\myX $, given by  
$\left(\myX  \right)_i = \sum_{l=1}^{\lenS} \left( \myS\right)_l a_{i-l+1} + v_i$, 
where $a_i$ is a deterministic known training sequence, and  $\{v_i\}_{i=1}^{\lenX}$ are samples from an i.i.d. zero-mean unit variance Gaussian noise process independent of $\myS$. 
In particular, the channel $\myS$ is modeled as an $\lenS =8$ tap zero-mean Gaussian vector with covariance matrix  $\CovMat{\myS}$, given by  $\big( \CovMat{\myS}\big)_{i,j} = e^{-|i-j|}$, $i,j \in \{1,2,\ldots,\lenS\}$, and  $a_i = \cos\left(\frac{2\pi i}{\lenX} \right)$ for $i >0$ and $a_i = 0$ otherwise. Since $\myS$ and $\myX$ are jointly Gaussian,  the \ac{mmse} estimate  is a linear function of $\myX$.

\begin{figure}
   	\centering
   	\begin{minipage}{0.45\textwidth}
   		\centering
   		\scalebox{0.45}{\includegraphics{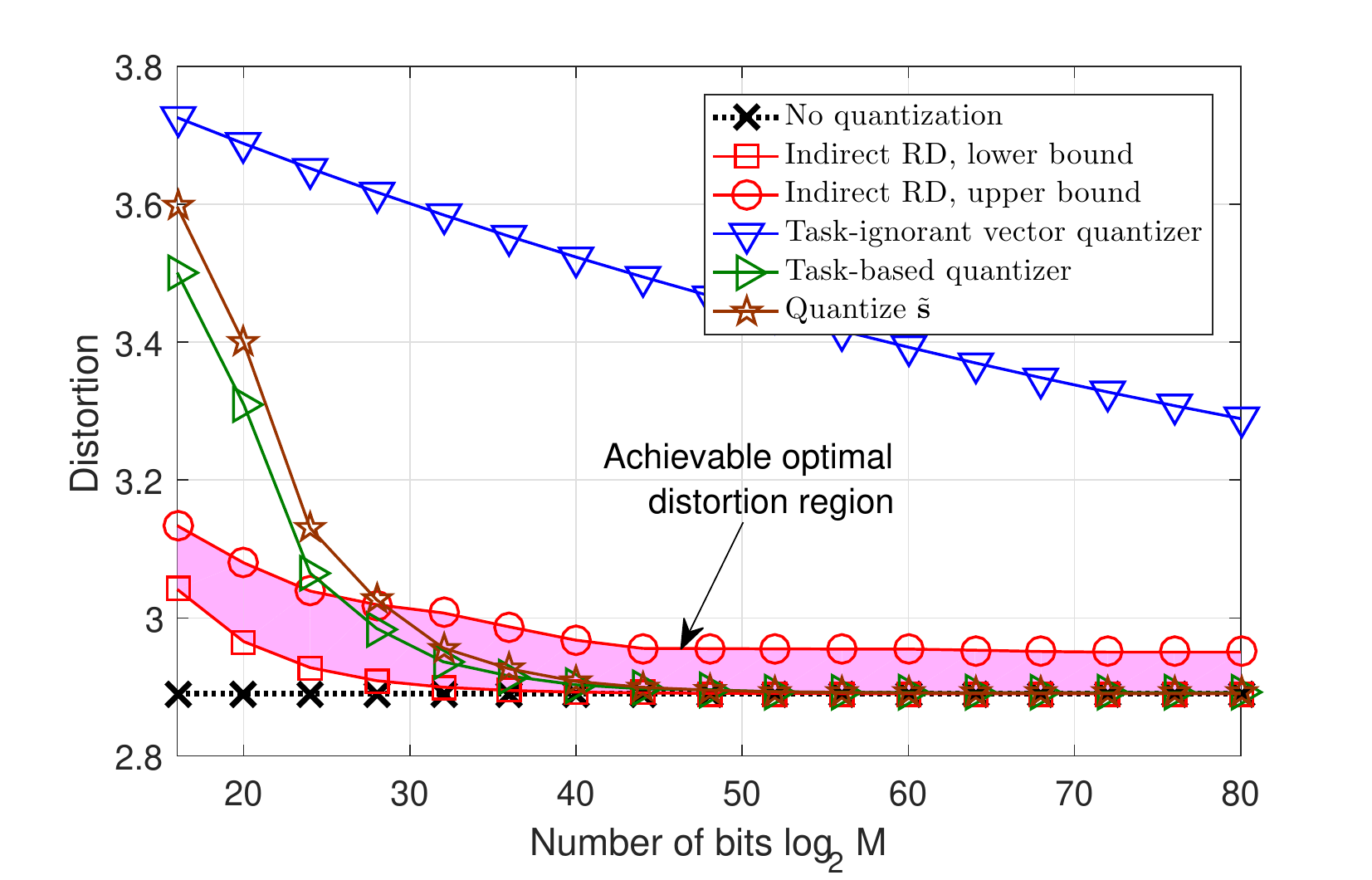}}
   		\vspace{-0.3cm}
   		\caption{\ac{isi} channel recovery.}
   		\label{fig:ChEst_K8}		
   	\end{minipage}
   	$\quad$
   	\begin{minipage}{0.47\textwidth}
   		\centering
   		\scalebox{0.45}{\includegraphics{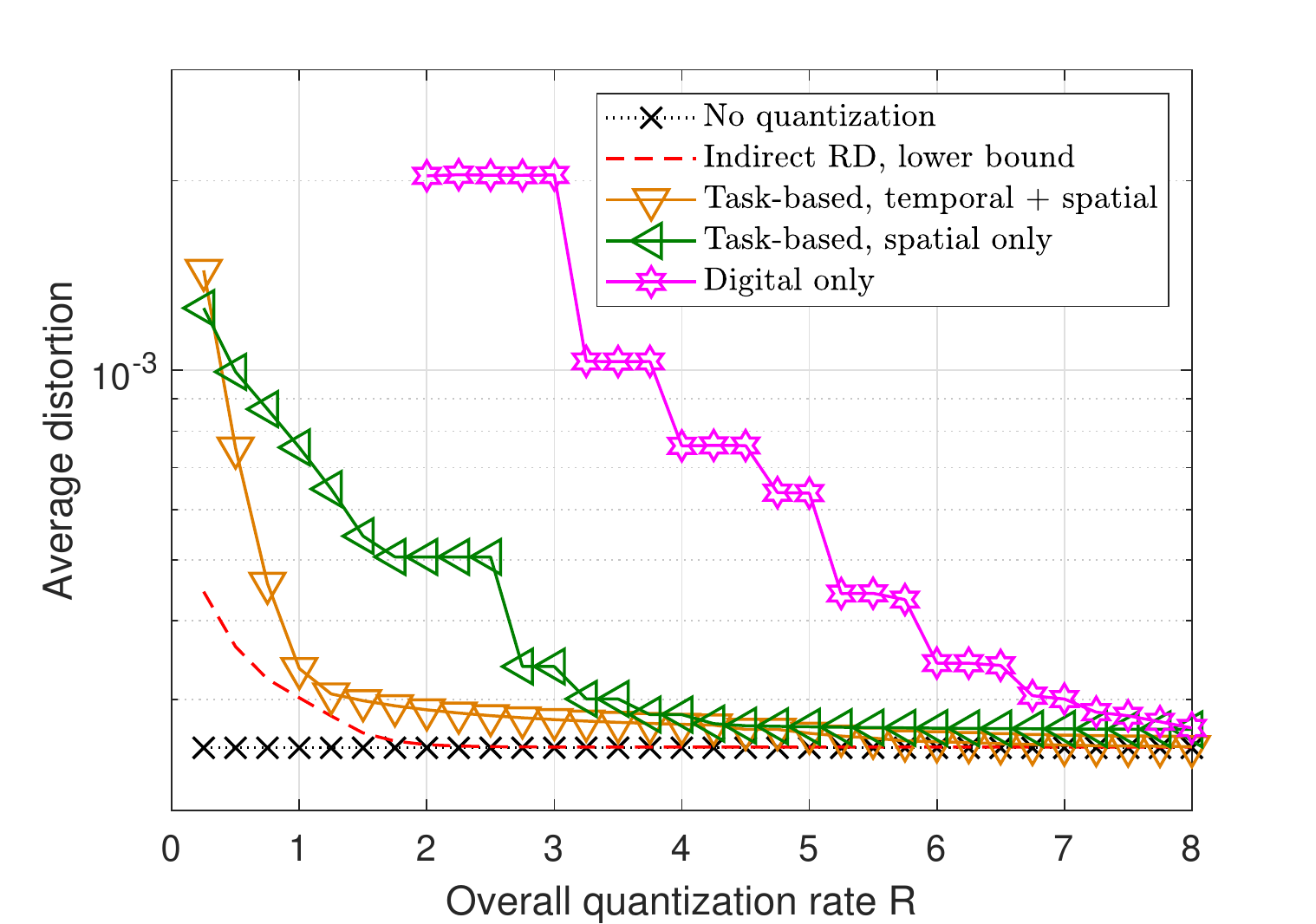}}
   		\vspace{-0.3cm}
   		\caption{\ac{mimo} channel recovery.}
   		\label{fig:DistVsRate_Corr3}
   	\end{minipage}
   \end{figure}

The \ac{mse} achievable by the task-based quantization system designed via Theorem \ref{thm:OptimalDes} operating with conventional non-dithered uniform quantizers is compared to the \ac{mse} in recovering the \ac{mmse} estimate in analog prior to quantization, i.e., setting $\myA = \LmmseMat$. We also numerically evaluate upper and lower bounds on the minimal \ac{mse} under quantization constraints, achievable via indirect rate-distortion theory by applying the rate-distortion optimal source code to $\tilde{\myS}$ (and thus given explicitly only in the limit $\lenS\rightarrow \infty$ \cite{kostina2016nonasymptotic}), computed via \cite[Prop. 1]{shlezinger2018hardware}. Finally, we evaluate the achievable \ac{mse} in applying a vector quantizer designed to accurately represent $\myX$, from which $\myS$ is estimated in digital, computed via \cite[Prop. 2]{shlezinger2018hardware}. The latter intuitively represents the vector quantization system one would design without prior knowledge of the task for which $\myX$ is acquired, and is thus referred to as task-ignorant vector quantizer. The \ac{mse} values are depicted in  Fig.  \ref{fig:ChEst_K8}. 

Observing Fig. \ref{fig:ChEst_K8}, 
we note that the task-based quantizer substantially outperforms task-ignorant vector quantization, and approaches the optimal performance as $M$ increases. In particular, when each scalar quantizer uses at least five bits, i.e., $\log M \ge 5\lenS$, the quantization error becomes negligible and the overall distortion is effectively the minimum achievable estimation error, i.e., the \ac{mmse}. 
Furthermore, we note that task-based quantization outperforms recovering $\tilde{\myS}$ in analog, and the gain is most notable  at small values of $M$.  These results demonstrate that by accounting for the presence of a task via joint optimization of the analog combiner, quantization rule, and digital processing, one can approach the optimal performance, dictated by indirect rate-distortion theory, using standard uniform \acp{adc} commonly used in digital signal processing systems. 

 To evaluate the performance of task-based quantization in massive \ac{mimo} systems, we consider a the recovery of multi-cell \ac{mimo} channel based on the setup detailed in \cite[Sec. V]{shlezinger2018asymptotic}. Here, the system consists of $7$ cells with $10$ single-antenna \acp{ut} in each cell, and the receiver, that is equipped with $100$ antennas, estimates its intra-cell   $100 \times 10$ channel matrix from the channel output, which is corrupted by intercell interference and Gaussian noise with variance of $10^{-3}$. The \acp{ut} are uniformly distributed in a hexagonal cell of radius $400$ m, following the model in \cite{marzetta2010noncooperative},  with receive side correlation dictated by Jakes model with $0.4$ wavelength element spacing \cite{jakes1994microwave}. Estimation is carried out based on $40$ pilot symbols determined by the first $10$ rows of the $40\times40$ \ac{dft} matrix.  
 
 The average \ac{mse} of the proposed task-based quantizer compared to the indirect rate-distortion bound and the \ac{mmse} achievable without quantization constraints is depicted in Fig. \ref{fig:DistVsRate_Corr3}. The input vector $\myX$ here represents the channel outputs corresponding to all transmitted pilot symbols, and thus the system designed via Theorem \ref{thm:OptimalDes} combines samples received at different time instances, which may be difficult to implement in practice. Therefore, we also depict in  Fig. \ref{fig:DistVsRate_Corr3} the \ac{mse} when the analog processing is restricted to combine only samples received at the same time instance using the same linear mapping, i.e., spatial only combining, obtained using \cite[Prop. 4]{shlezinger2018asymptotic}. Finally, we  depict the \ac{mse} without analog combining, i.e., a digital only receiver, in which the digital processing is based on the linear \ac{mmse} channel estimator from quantized measurements, and thus consists a bound on the performance achievable using approximations of the linear \ac{mmse} estimator, such the channel estimator  proposed in \cite{jacobsson2017throughput}.  
 
 Observing Fig. \ref{fig:DistVsRate_Corr3} we note that, similarly to the \ac{isi} channel in Fig. \ref{fig:ChEst_K8}, the \ac{mse} achievable using task-based quantization is within a very small gap from the indirect rate-distortion curve for quantization rates larger than $\Rate = 1.5$. 
 The task-based quantizer with spatial combining is capable of achieving near-optimal performance for $\Rate > 3$, due to its  ability to exploit the spatial correlation. It is also observed that the average \ac{mse} of estimating the channel only in the digital domain is notably higher compared to task-based quantization, which jointly operates in both analog and digital while tuning the quantization rule accordingly, demonstrating the gains of task-based quantization over digital-only designs.

\vspace{-0.2cm}
\subsection{Quadratic Estimation Tasks}
\label{subsec:TaskQuadratic}
\vspace{-0.2cm}
	In the previous section we showed that allowing the analog mapping to reduce  dimensionality and rotate the quantized signal can contribute to the overall recovery performance by balancing estimation and quantization errors. However, this analysis was carried out only for scenarios in which $\tilde{\myS}$ is a linear function of $\myX$, resulting in   $\E\big\{\myS | \myZ \big\}$ being a linear function of the input to the quantizers $\myZ$.  In many scenarios of interest, such as covariance estimation \cite{rodrigues2017rate} and \ac{doa} recovery \cite{yu2016doa} from quantized measurements, the desired information can be extracted from a quadratic function of the measurements, i.e., functions $\{\myX^T \myMat{C}_{i} \myX\}_{i=1}^{\lenS}$, where each $\myMat{C}_{i} \in \mySet{R}^{\lenX \times \lenX}$ is symmetric. 
	
	Here, we show how the analysis of the previous section can be applied for designing task-based quantizers for the task of recovering non-linear functions of $\myX$ under quantization constraints, focusing on quadratic functions and Gaussian inputs. 
%
%
Our strategy is based on identifying a family of analog mappings $h(\cdot)$ for which $\myZ$ corresponds to the scenario studied in Section \ref{subsec:TaskLinear}. To that aim, we use \ac{pic}-based analysis \cite{du2017principal}, which provides a decomposition of the statistical relationship between two \acp{rv}, that is directly related to \ac{mmse} estimation. 
In particular, for a pair of \acp{rv} $(x,y)$, the principal inertia functions $\{f_i(\cdot)\}$ and $\{g_i(\cdot)\}$ formulate an orthonormal basis spanning the Hilbert space of functions of $x$ and $y$, respectively, which diagonalize \ac{mmse} estimation, i.e., there exists a set of scalar coefficients $\{\rho_i\}$ such that $\E\{f_i(x)|y\} = \rho_i g+i(y)$ and $\E\{g_i(y)|x\} = \rho_i f_i(x)$. The benefit of using \acp{pic} in our context is their ability to decompose functions of the observations in a manner which reflects on the structure of the \ac{mmse} estimate. In particular, here we use this tool to identify a transformation of the input $\myX$ under which recovering quadratic functions of it is converted to a linear manipulation.
Defining $\bar{\myX} \triangleq {\rm vec}(\myX \myX^T)$, this results in the following theorem \cite[Thm. 1]{salamatian2019task}:
\begin{theorem}
	\label{thm:PICQuad}
	For any $\lenZ \times \lenX^2$ matrix $\myMat{A}$ with $\lenZ \le \lenX^2$, the \ac{mmse} estimate of $f(\myX) = \myX^T\myMat{C}\myX$ from  the vector $\myZ = \myMat{A}(\bar{\myX} - \E\{\bar{\myX}\})$ can be written as
	\begin{equation}
	\label{eqn:PICQuad}
	\E\big\{f(\myX) | \myZ  \big\} = \myVec{d}^T \myZ + \E\{f(\myX)\},
	\end{equation}
	for some $\lenZ \times 1$ vector $\myVec{d}$, which depends on $\myMat{C}$, $\myMat{A}$, and the covariance of $\myX$. 
\end{theorem}

\begin{figure}
	\centering
	\includegraphics[width = 0.8\columnwidth]{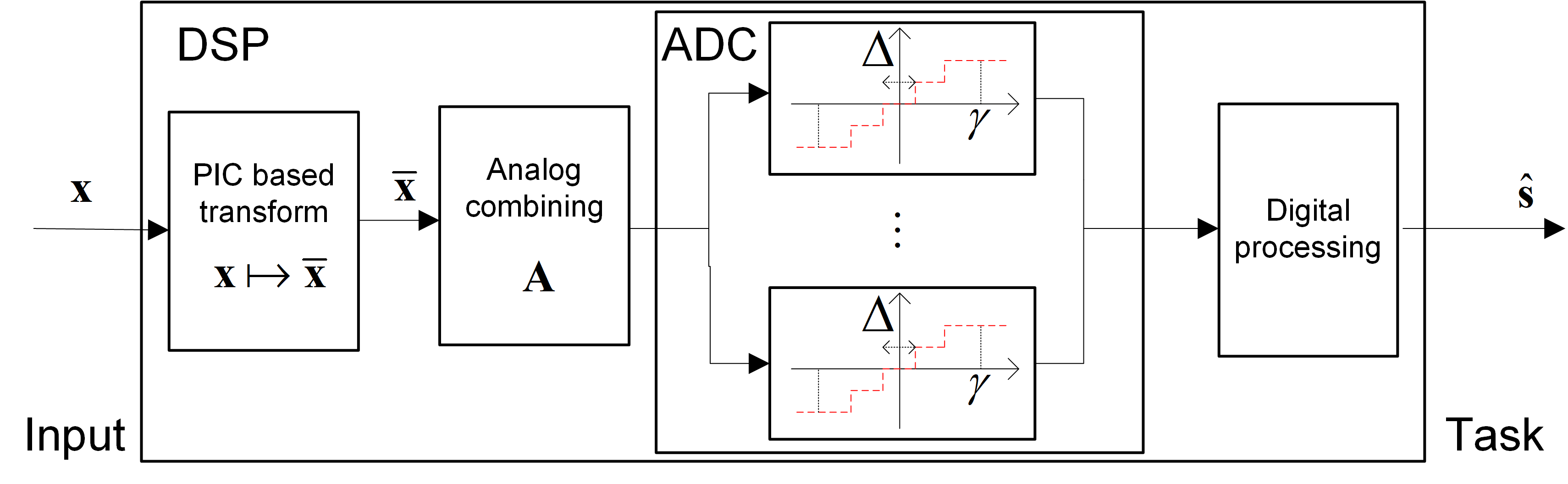}	
	\vspace{-0.2cm}	
	\caption{Quantization System of Fig. \ref{fig:SystemModel} with quadratic analog mapping.} 
	\label{fig:system2}
\end{figure}

Theorem \ref{thm:PICQuad} implies that the task-based quantization system design guidelines proposed in Theorem \ref{thm:OptimalDes} can be utilized to facilitate the recovery of quadratic functions from quantized measurements by applying analog mappings of the form $\myZ = \myMat{A}h(\myX) = \myMat{A}(\bar{\myX} - \E\{\bar{\myX}\})$. Here the matrix $\myMat{A} \in \mySet{R}^{\lenZ \times \lenX^2}$ encapsulates the ability to reduce  the dimensionality and to rotate the quantized vector, and can be designed via Theorem \ref{thm:OptimalDes} by replacing the input $\myX$ with $\bar{\myX} - \E\{\bar{\myX}\}$.   
The resulting quantization system is depicted in Fig. \ref{fig:system2}.

Although Theorem~\ref{thm:PICQuad} specifically considers functionals $f(\myX)$ of a quadratic form,   analogous schemes could be constructed for broader classes of functions. The main feature of Theorem~\ref{thm:PICQuad} is the ability to represent $\E\{f(\myX)|\myZ\}$ either exactly, or possibly approximately, as a linear function of $\myZ = \myA h(\myX)$ for some transformation $h(\cdot)$. Once the analog mapping satisfies this request, Theorem \ref{thm:OptimalDes} can be applied to optimize the overall recovery accuracy of the quantization system. 
Formulated in terms of \acp{pic}, the choice of  $h(\cdot)$ imposes  structure on the joint distribution $(\myX,\myZ)$. Consequently, when the task is to recover a function $f(\myX)$ which can be decomposed using \acp{pic} as $f(\myX) = \sum \alpha_i f_i(\myX)$, any analog processing which results in $\myZ$ such that
\vspace{-0.2cm}
\begin{align}
\E\{f(\myX)|\myZ\} \approx \sum_{i = 1}^l \alpha_i \rho_i (\myZ)_i + \E\{f(\myX)\}, 
\vspace{-0.2cm}
\end{align}
 would allow to design the analog pre-quantization step using existing tools derived for setups in which the \ac{mmse} estimate is linear. This implies that when recovering  some function $f(\myX)$, the structure of the analog mapping should be designed as to yield linear basis functions $g_i(\myZ)$, allowing the resulting system to be optimized using Theorem \ref{thm:OptimalDes}.

\begin{figure}
	\centering
	\scalebox{0.45}{\includegraphics{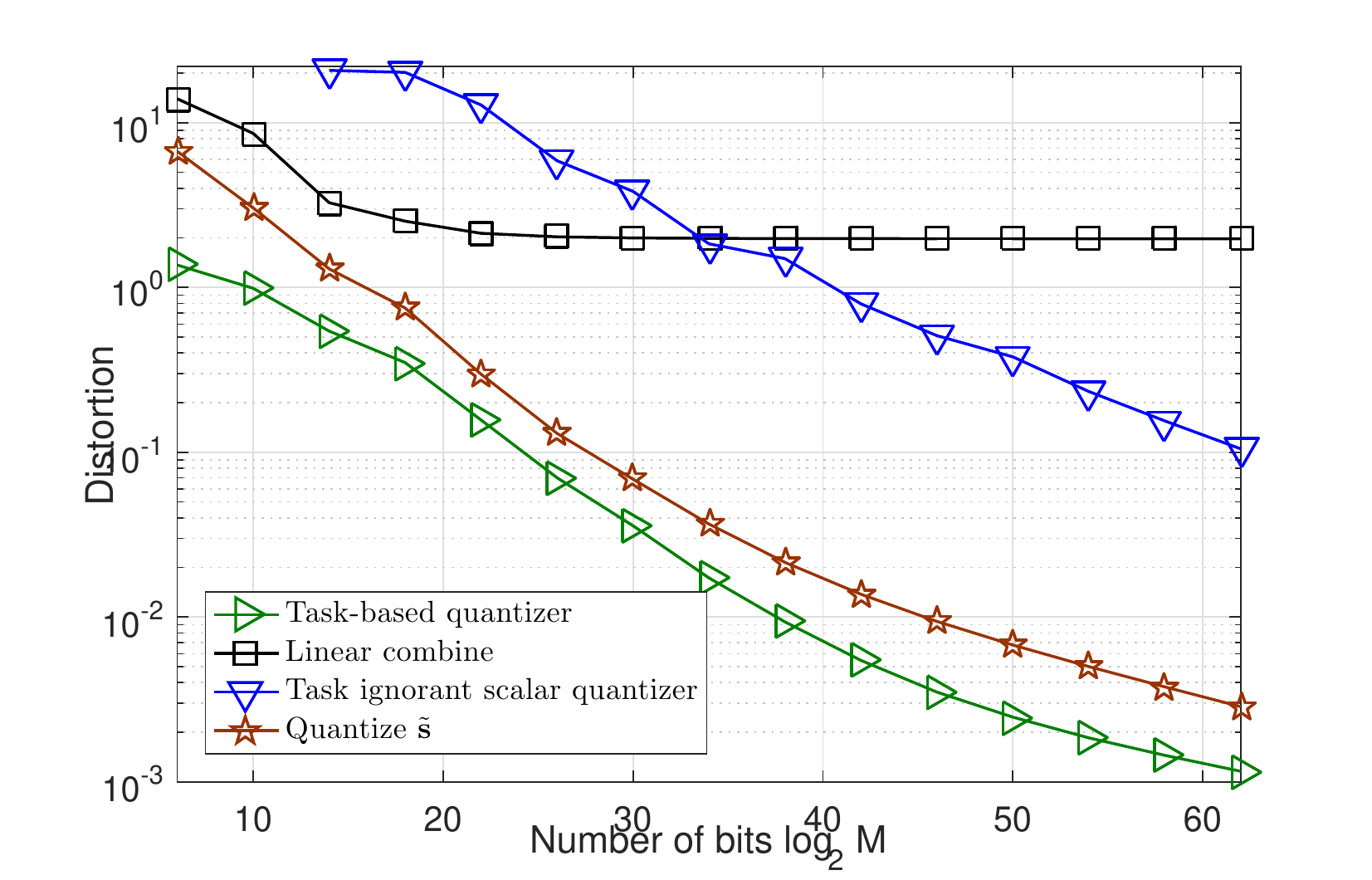}}
	\vspace{-0.6cm}
	\caption{Empirical covariance recovery.
	}
	\label{fig:Covariance1}
\end{figure}

To demonstrate the ability of the proposed design to yield accurate task-based quantizers, we simulate an empirical covariance estimation scenario. Here, the input is given by $\myX = [\myVec{v}_1^T,\ldots \myVec{v}_4^T]^T$, where $\{\myVec{v}_i\}_{i=1}^{4}$ are i.i.d. $3 \times 1$ zero-mean  Gaussian random vectors, i.e., $\lenX =12$. The entries of the covariance matrix of $\myVec{v}_i$, denoted $\CovMat{\myVec{v}}$, are $\left( \CovMat{\myVec{v}}\right)_{i,j} = e^{-|i-j|}$. The parameter of interest is the $3 \times 3$ empirical covariance matrix $\frac{1}{4}\sum_{i=1}^{4} \myVec{v}_i\myVec{v}_i^T$, which is completely determined by its upper triangular matrix, stacked as the desired vector $\tilde{\myS}$, thus $\lenS = 6$. For the considered scenario,  we evaluate the \ac{mse} achievable by the task-based quanziation system of Fig. \ref{fig:system2} where the analog combiner, quantization support, and digital processing are obtained via Theorem \ref{thm:OptimalDes}. The task-based quantizer is compared to recovering the empirical covariance in analog, as well as to directly quantizing $\myX$, i.e., a task-ignorant scalar quantizer, and a hybrid system utilizing linear analog combiners based on \cite[Sec. V]{shlezinger2018hardware}.
	For all the above systems, in order to avoid overloading the quantizers, the support is set to $\myEta$ times the maximal sum of the standard deviation and  absolute mean value of the entries of the input to the \ac{adc}, where we let $\eta$ increase linearly with the number of bits in the range $[3,6.5]$. 
The achievable \acp{mse} versus the number of bits are depicted in Fig. \ref{fig:Covariance1}. 
Observing Fig. \ref{fig:Covariance1}, we note that the task-based quantizer, which is designed to balance the quantization and estimation errors, achieves the best \ac{mse} performance. 
Quantizing $\tilde{\myS}$ directly results in notable quantization errors when operating with a  small number of bits, due to the need to set the support to a relatively large value resulting in coarse quantization. This demonstrates how the task-based quantization design proposed in Section \ref{subsec:TaskLinear} for linear tasks can be extended to apply for recovering non-linear functions.

\vspace{-0.2cm}
\section{Deep Task-Based Quantization}
\label{sec:TaskDeep}
\vspace{-0.2cm}
In Section \ref{sec:Task} we designed hybrid analog-digital acquisition systems, which consist of analog combining, scalar quantization, and digital processing, to accurately recover some underlying information embedded in the observed analog signal. The systems proposed in Section \ref{sec:Task} are { model-aware}, requiring accurate knowledge of the statistical relationship between the observations and the task, i.e., $\Pdf{\myX | \myS}$. 
Two notable challenges are associated with such model-aware designs: 
$1)$ Accurate knowledge of the statistical model $\Pdf{\myX | \myS}$ may be unavailable in practice;
$2)$ Even when $\Pdf{\myX | \myS}$ is perfectly known, analytically tractable characterizations  are obtained only for tasks of relatively simple form, e.g., linear and quadratic functions, under the model assumptions \ref{itm:A1}-\ref{itm:A2}. This limits the design to estimation tasks  \ref{itm:A1}, does not explore arbitrary quantization rules \ref{itm:A2}, and may not lead to analytically tractable systems when operating under complex statistical relationships.

An alternative approach to inferring the quantization system from the model, is to learn it from a set of training samples in a data-driven fashion. In particular, by utilizing \ac{ml} methods, one can  implement task-based quantizers without the need to explicitly know the underlying model and to analytically derive the proper quantization rule. Furthermore, when the parameters of the hybrid analog-digital system are learned from data and not specified analytically, the quantization mapping can be optimized along with the system parameters instead of fixing a uniform rule as in \eqref{eqn:UniQuant}. Finally, additional families of tasks, such as classification, can be considered by properly setting the loss function utilized in the learning process.  

In this section we present a generic \acp{dnn} architecture which utilizes \ac{ml} for task-based quantization with scalar \acp{adc}, referred to as {\em deep task-based quantization} \cite{shlezinger2019deep}. We begin with the system architecture in Section \ref{subsec:DeepModel}, after which we present how the quantization mapping is learned in Section~\ref{subsec:DeepQuant}. We provide  numerical results along with a discussion in Section~\ref{subsec:DeepSims}.

\vspace{-0.2cm}
\subsection{System Architecture}
\label{subsec:DeepModel}
\begin{figure}
	\centering
	\includefig{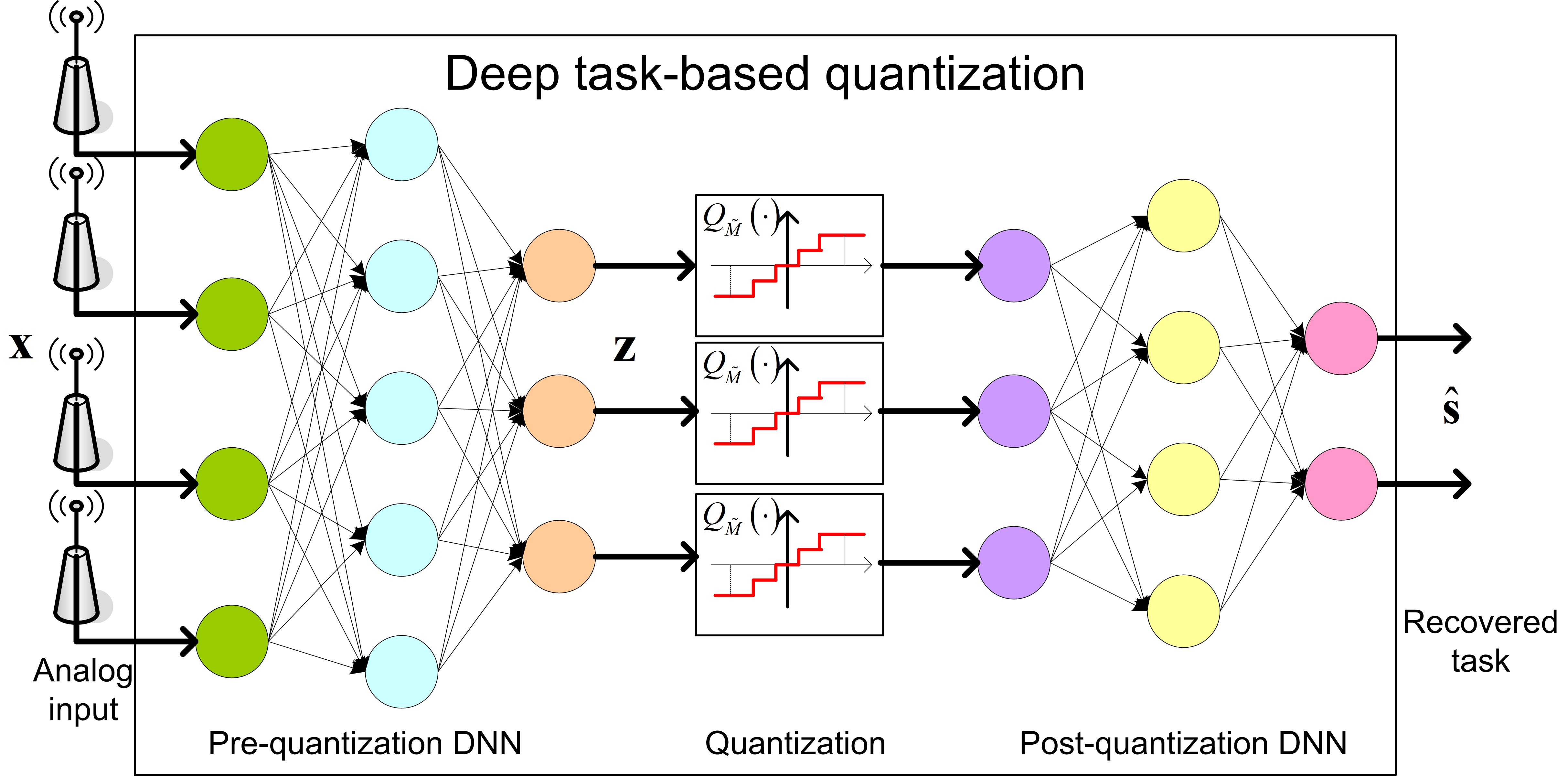}
	\caption{Deep task-based quantization system architecture.} 
	\label{fig:DNNModel}
\end{figure}

\vspace{-0.2cm}
Deep task-based quantization operates in a data-driven manner, learning the analog transformation, quantization mapping, and digital processing, from a training data set, consisting of $t$ independent realizations of $\myVec{s}$ and $\myVec{x}$, denoted $\{ \myVec{s}^{(i)}, \myVec{x}^{(i)}\}_{i=1}^{t}$.
In general, the training samples may be taken from a set of joint distributions, and not only from the true (unknown) joint distribution of $\myVec{s}$ and $\myVec{x}$.
Here, the analog pre-quantization mapping and the digital post-quantization processing are parameterized as layers of a \ac{dnn}, as illustrated in Fig.~\ref{fig:DNNModel}.  By doing so, the overall task-based quantization system, including the analog combining, quantization rule, and digital processing, can be trained from data in an end-to-end manner using e.g., \ac{sgd}. While the proposed system focuses only on the quantization aspect of \acp{adc}, the resulting design approach can be extended to account also for sampling in addition to quantization, as considered in \cite{shlezinger2020learning}.

In the proposed architecture, the scalar \ac{adc}, which implements the continuous-to-discrete mapping, is modeled as an activation function between  two intermediate layers, interfacing the analog processing and the digital part. The trainable parameters of this activation function determine the quantization rule, allowing it to be learned during training.
The \ac{dnn} structure cannot contain any skip connections between the multiple layers prior to quantization (analog domain) and those after quantization (digital domain), representing the fact that all analog values must be first quantized before processed in digital. The pre and post quantization networks are henceforth referred to as the {\em analog \ac{dnn}} and the {\em digital \ac{dnn}}, respectively. 
The system input is the observed  $\myVec{x}$, and we use $\myVec{\theta}$ to denote the network parameters. Two families of tasks are considered:
\begin{itemize}
	\item {\bf Estimation}: Here, the system should learn to recover a set of $\lenS$ unknown parameters taking values on a continuous set, i.e., $\mySet{S} = \mySet{R}$. By letting  $\NetMap(\cdot)$ denote the mapping implemented by the overall system, the output is  given by the $\lenS \times 1$ vector $\hat{\myVec{s}} = \NetMap(\myVec{x})$, which is used as a representation of   $\myVec{s}$. The loss function is the empirical \ac{mse}:
	\vspace{-0.2cm}
	\begin{equation}
	\label{eqn:LossFunc}
	\mathcal{L}(\myVec{\theta})  =\frac{1}{t}\sum_{j=1}^{t}\left\Vert \myS^{\left(j\right)}-\NetMap\Big(\myVec{x}^{(j)}\Big)\right\Vert _{2}^{2}.
	\vspace{-0.2cm}
	\end{equation} 
	\item {\bf Classification}: In such tasks, the system should decide between a finite number of options. Here, $\mySet{S}$ is a finite set, and we use $|\mySet{S}|$ to denote its cardinality. The last layer of the digital \ac{dnn} is a softmax layer, and thus the network mapping  $\NetMap(\cdot)$ is a $|\mySet{S}|^\lenS \times 1$ vector, whose entries represent the conditional probability for each different value of $\myVec{s}$ given the input $\myVec{x}$. By letting $\NetMap(\myVec{x} ; \myVec{\alpha})$ be the output value corresponding to $\myVec{\alpha} \in \mySet{S}^{\lenS}$, the decision is selected as the most probable one, i.e., $\hat{\myVec{s}} = \arg\max_{\myVec{\alpha} \in \mySet{S}^{\lenS}}  \NetMap(\myVec{x} ; \myVec{\alpha})$.  	 The loss function is the empirical cross-entropy:
	\vspace{-0.2cm}
	\begin{equation}
	\label{eqn:LossFuncCE}
	\mathcal{L}(\myVec{\theta})  =\frac{1}{t}\sum_{j=1}^{t} -\log  \NetMap\Big(\myVec{x}^{(j)} ; \myVec{s}^{(j)} \Big).
	\vspace{-0.2cm}
	\end{equation} 
\end{itemize} 

\vspace{-0.2cm}
\subsection{Learned Quantization Mappings}
\label{subsec:DeepQuant}
\vspace{-0.2cm}
The proposed architecture implements scalar quantization as an intermediate activation in a joint analog-digital hybrid \ac{dnn}. This layer converts its continuous-amplitude input into a discrete quantity.  The non-differentiable nature of such continuous-to-discrete mappings induces a  challenge in applying \ac{sgd} for optimizing the  network parameters.  
In particular, quantization activation, which can be modeled as a superposition of step functions determining the continuous regions jointly mapped into a single value, nullifies the gradient of the cost function. Thus,  straight-forward application of \ac{sgd} with back-propagation fails to properly set the pre-quantization network.

This challenge can be tackled by approximating the non-differentiable quantization mapping by a differentiable one, as proposed in \cite{agustsson2017soft}. This is achieved by replacing the continuous-to-discrete transformation with a non-linear activation function which has approximately the same behavior as the quantizer. Specifically, we use a sum of shifted hyperbolic tangents, which are known to closely resemble step functions in the presence of large magnitude inputs. The resulting scalar quantization mapping is given by:
\vspace{-0.1cm}
\begin{equation}
\label{eqn:tanh}
\tilde{q}(z)=\sum_{i=1}^{\tilde{M}-1}a_{i}\tanh\left(c_{i}\cdot   z-b_{i} \right),
\vspace{-0.1cm}
\end{equation}
where $\{a_i, b_i, c_i \}$ are  real-valued parameters. When the parameters $\{c_i\}$ increase, the corresponding hyperbolic tangents approach step functions. 

\begin{figure}
	\centering
		\scalebox{0.45}{\includegraphics{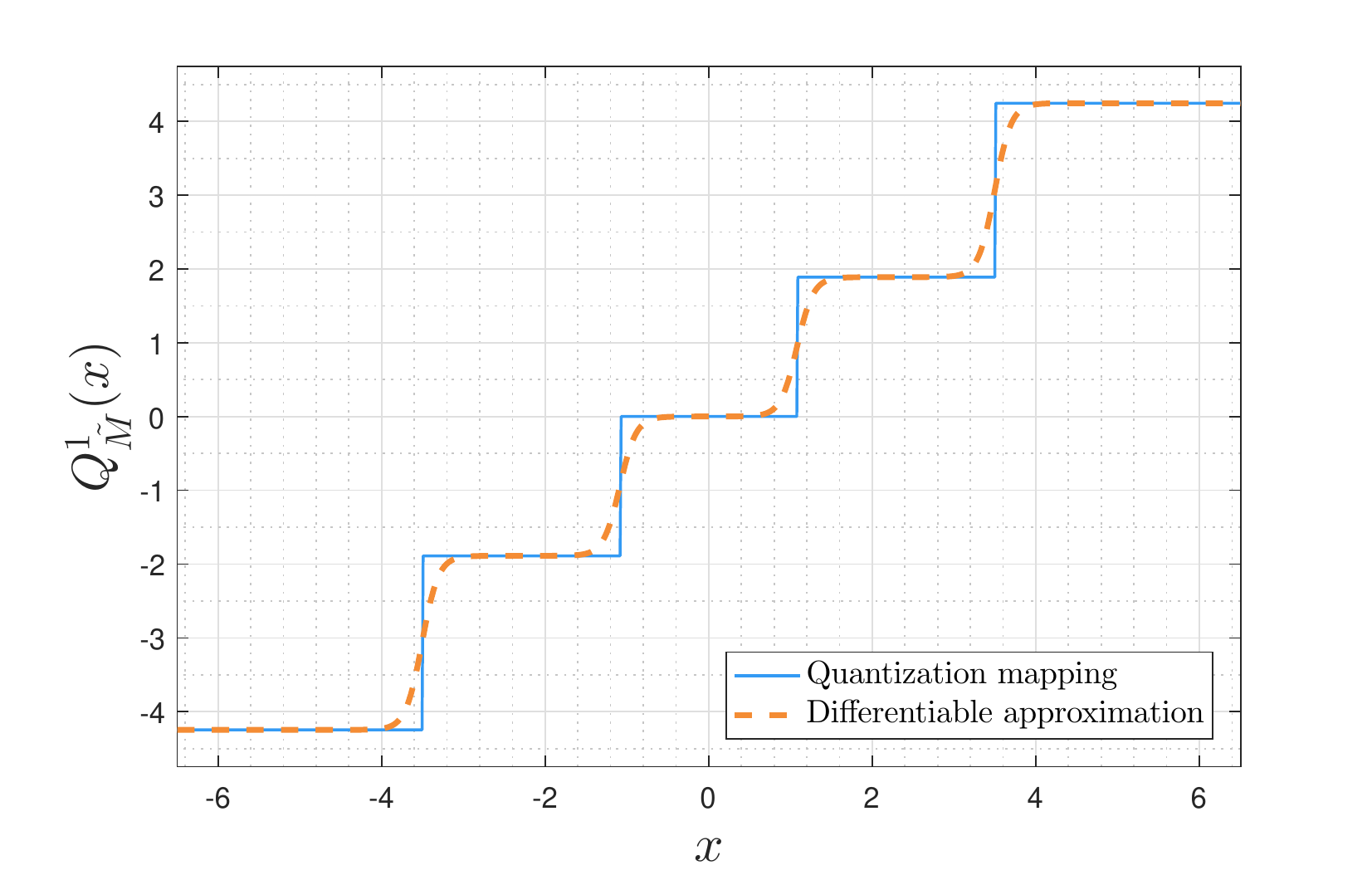}} 
		\vspace{-0.4cm}
	\caption{Differentiable approximation of the quantization rule illustration.} 
	\label{fig:Tanh_Aproximation}
\end{figure}

In addition to learning the weights of the analog and digital \acp{dnn}, this  approach allows  to learn the quantization  function, and particularly, the best suitable constants $\{a_{i}\}$ and $\{b_{i}\}$. These tunable parameters are later used to determine the decision regions of the scalar quantizer, where the set $\{b_{i}\}$ is used for the decision regions limits while  $\{a_{i}\}$ determines the corresponding discrete values assigned to each decision region.  
The parameters $\{c_i\}$, which essentially control the resemblance of \eqref{eqn:tanh} to an actual continuous-to-discrete mapping, do not reflect on the quantization rule, and are thus not learned from training. 
The proposed optimization is achieved by including the parameters  $\{a_i, b_{i}\}$  as part of the network trainable parameters $\myVec{\theta}$. 
Due to the differentiability of  \eqref{eqn:tanh}, one can now apply standard \ac{sgd} with back-propagation to optimize the overall network, including the analog and digital \acp{dnn} as well as the quantization rule, in an end-to-end manner.
Once training is concluded,  the learned $\tilde{q}(\cdot)$ activation \eqref{eqn:tanh} is replaced with a scalar quantization mapping dictated by the tunable parameters $\{a_i, b_i \}$.  An illustration of how the differentiable mapping \eqref{eqn:tanh} is converted into a continuous-to-discrete quantization rule  is depicted in Fig. \ref{fig:Tanh_Aproximation}. The dashed smooth curve in Fig. \ref{fig:Tanh_Aproximation} represents the differentiable function after training is concluded, and the straight curve is the resulting scalar quantizer.

\vspace{-0.2cm}
\subsection{Numerical Results}
\label{subsec:DeepSims}
\vspace{-0.2cm}
We next numerically demonstrate the achievable performance of deep task-based quantization. In the following, we model the relationship between the observed $\myX$ and the task $\myS$ as
\begin{equation}
\label{eqn:Received1}
\myX = \myMat{H} \myS + \myVec{w},
\end{equation} 
for some fixed $\myMat{H} \in \mySet{R}^{\lenX \times \lenS}$, where $\myVec{w}  \in \mySet{R}^\lenX$ is a zero-mean Gaussian vector with i.i.d. entries of variance $\sigma_w^2 > 0$. 

\begin{figure}
	\centering
	\begin{minipage}{0.45\textwidth}
		\centering
		\scalebox{0.45}{\includegraphics{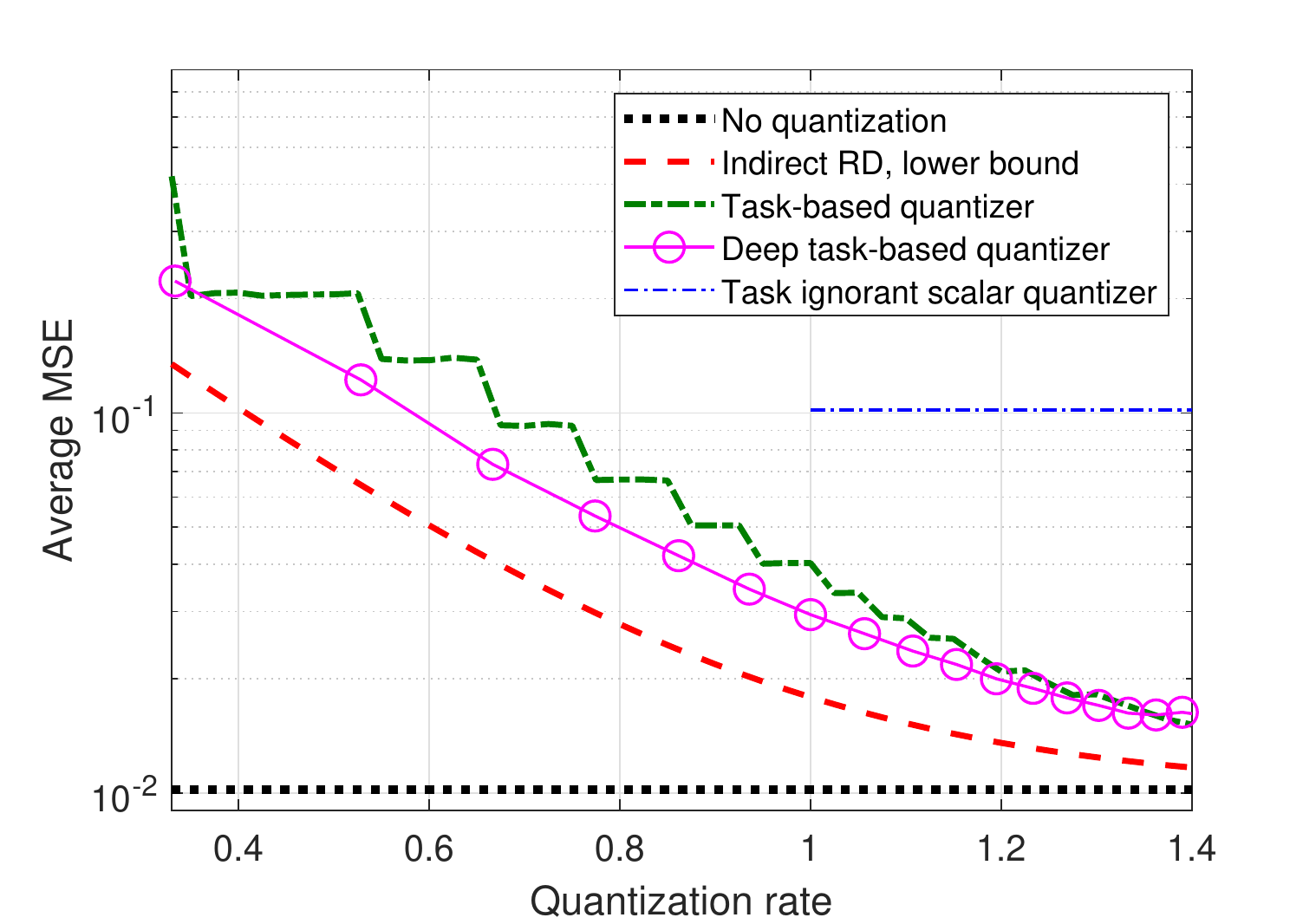}}
		\vspace{-0.3cm}
		\caption{Estimation task.}
		\label{fig:Deep_ChEst1}		
	\end{minipage}
	$\quad$
	\begin{minipage}{0.45\textwidth}
		\centering
		\scalebox{0.45}{\includegraphics{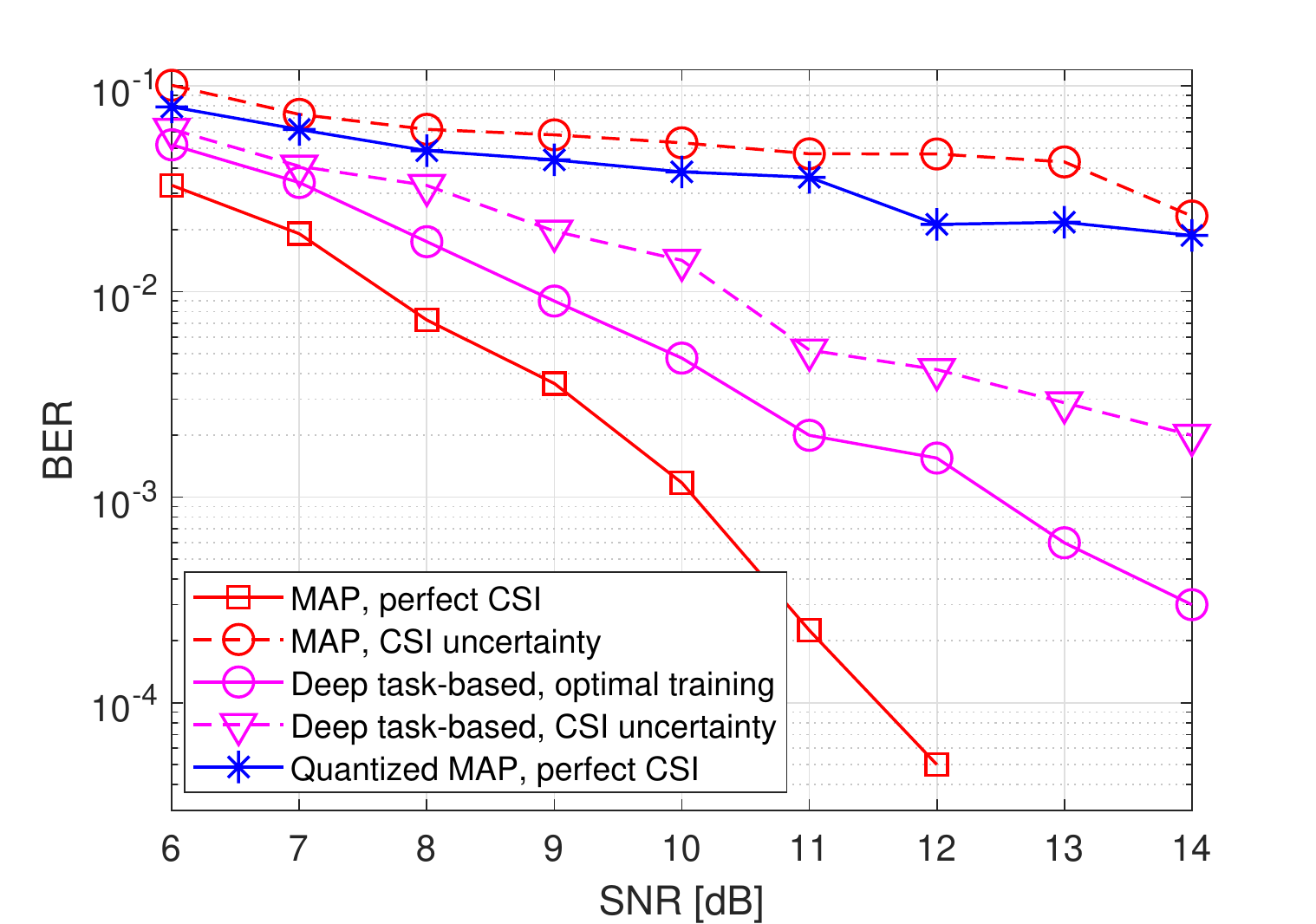}}
		\vspace{-0.3cm}
		\caption{Classification task.}
		\label{fig:Deep_SymRec1}
	\end{minipage}
\end{figure}

We begin with an estimation task for which we can compare the data-driven task-based system to its model-aware counterpart detailed in Section~\ref{subsec:TaskLinear}. Here, we set $\sigma_w^2 = 0.25$, $\lenX = 120$, $\lenS = 40$, while  $\myS$ is a zero-mean Gaussian vector with i.i.d. unit variance entries. The matrix $\myMat{H}$ is set to 
\begin{equation*}
\myMat{H} = \left[\begin{array}{cc}
{\rm Re}\big(\myMat{\Phi} \otimes \myMat{I}_{10}\big) &  
{\rm Im}\big(\myMat{\Phi} \otimes \myMat{I}_{10}\big) \\
-{\rm Im}\big(\myMat{\Phi} \otimes \myMat{I}_{10}\big) &
{\rm Re}\big(\myMat{\Phi} \otimes \myMat{I}_{10}\big)
\end{array}\right],
\end{equation*} 
where $\myMat{\Phi}$ is the first $4$ columns of the $12\times 12$ \ac{dft} matrix. This setting represents channel estimation in Rayleigh fading \ac{mimo} channels using orthogonal pilots \cite[Sec. IV]{shlezinger2019deep}. 
In Fig. \ref{fig:Deep_ChEst1} 
we numerically evaluate the average \ac{mse} versus the quantization rate $\Rate$ of deep task-based quantization compared to the fundamental performance limit dictated by indirect rate-distortion theory, as well as to the performance of the model-aware task-based quantizer discussed in Section \ref{sec:Task}.  To guarantee fair comparison with the model-aware system we set the pre and post quantization \acp{dnn} to consist of linear layers. 
Following \cite[Prop. 2]{shlezinger2018hardware}, we set the number of scalar quanizers to $\lenZ = \lenS$ for both task-based quantizers.
The data-driven system is trained using $t=2^{15}$ labeled pairs, and all systems are tested using $2^{10}$ test samples. We also depict in Fig. \ref{fig:Deep_ChEst1} the average \ac{mse} of a task-ignorant system in which estimation is carried out only in the digital domain, using the method for channel estimation from quantized measurements proposed in \cite{jacobsson2017throughput}.

Observing Fig. \ref{fig:Deep_ChEst1}, we note that the fact that data-driven quantizer is not restricted to uniform quantizers allows it to outperform the model-aware system of Section \ref{sec:Task} especially in lower quantization rates. Furthermore, the performance of both task-based quantizers is within a relatively small gap of the fundamental performance limits. 
These results demonstrate the ability of deep task-based quantization to implement a feasible and optimal-approaching quantization system in a data-driven fashion. 

Next, we consider a classification task. Again, the observations $\myX$ are related to the task vector $\myS$ via \eqref{eqn:Received1}. However, here the entries of $\myS$ are i.i.d. uniformly distributed over $\mySet{S} = \{-1, 1\}$ representing, e.g., symbol detection in \ac{mimo} communications. In particular, we use $\lenX = 12$, $\lenS = 4$, and set the entries of $\myMat{H}$ to $(\myMat{H})_{i,j} = e^{-|i-j|}$.  
For the deep task-based quantizer we use two fully-connected layers in analog and two fully-connected layers in digital. As this is a classification task, the output layer is a softmax function with $2^{\lenS}$ probabilities, and the overall network is trained to minimize the cross-entropy loss \eqref{eqn:LossFuncCE} using $t=5000$ labeled samples. Unlike the estimation task for which the number of quantizers $\lenZ$ can be set according to the analytical results in \cite{shlezinger2018hardware}, here this value was determined based on empirical evaluations. In particular, we use $\lenZ = \lfloor  \lenS \Rate \rfloor$, resulting in each scalar quantizer using at least $\lenX / \lenS = 3$ bits in the hybrid system.

 The numerically computed \ac{ber} averaged over $20000$ trials versus the \ac{snr} defined as $1/\sigma_w^2$ of the deep task-based quantizer with quantization rate $\Rate = 1$ is depicted in Fig. \ref{fig:Deep_SymRec1} compared to the \ac{map} rule operating for recovering $\myS$ from $\myX$, i.e., without quantization constraints, as well as the \ac{map} rule for recovering $\myS$ from a uniformly quantized $\myX$ with rate $\Rate = 1$, representing a task-ignorant digital only system. It is noted that the \ac{map} detectors require prior knowledge of $\myMat{H}$ or $\sigma_w^2$, while the data-driven quantizer is invariant of the underlying model and learns its mapping from training. In order to study the resiliency of  deep task-based quantization to inaccurate training, we also compute the \ac{ber} under \ac{csi} uncertainty, namely, when the training samples are randomized from a joint distribution of $\myVec{s},\myVec{x}$ in which the entries of the matrix $\myMat{H}$ in \eqref{eqn:Received1} are corrupted by additive i.i.d. Gaussian noise, whose variance is $20\%$ the magnitude of the corresponding entry. For comparison, we also evaluate the \ac{ber} of the \ac{map} rule with the same level of \ac{csi} uncertainty. 
 
 Observing Fig. \ref{fig:Deep_SymRec1}, we note that in the presence of accurate \ac{csi}, the  \ac{ber} of our deep task-based quantizer is comparable to that achievable using the \ac{map} rule operating without quantization constraints.  For comparison, the quantized \ac{map} rule, which operates only in the digital domain, achieves significantly worse \ac{ber} performance compared to the hybrid deep task-based quatizer, demonstating the benefit of  applying pre-quantization processing in the analog domain in order   to utilize more accurate quantization while keeping the semantic information required to carry out the task.  The results in Fig. \ref{fig:Deep_SymRec1}  also demonstrate the improved robustness of the data-driven system to inaccurate \ac{csi}. The performance of the model-based \ac{map} detector is very sensitive to \ac{csi} uncertainty, resulting in a notable increase in \ac{ber} due to the model mismatch. However, the performance of the  deep task-based quantizer trained under  \ac{csi} uncertainty is within an \ac{snr} gap of approximately $0.5-2$ dB from its  performance when trained using accurate \ac{csi}.  This demonstrates the gains of using \acp{dnn}  for overcoming the sensitivity of model-based approaches to inaccurate model knowledge.
 
\vspace{-0.2cm}
\section{Hardware Implementation for MIMO Receivers}
\label{sec:Practice}
\vspace{-0.2cm}
In the previous sections we presented the concept of task-based quantization, in which the components of a hybrid analog-digital system are jointly optimized to facilitate the recovery of some underlying information under bit constraints. We considered two complementary strategies for tuning task-based quantizers: a model-aware approach and a data-driven method. Here, we discuss how the systems designed using either of the aforementioned strategies can be realized, as well as which additional practical considerations must be taken into account and how they can be incorporated in the design. We focus here on task-based quantization for \ac{mimo} receivers, in which multiple signals are acquired for some task other than recovering them in digital, and where quantization constraints play an important role. 

Conventional \ac{mimo} receivers obtain their observations using a set of antennas, where each antenna is connected to a dedicated scalar \ac{adc}, typically implementing a uniform quantization mapping. Consequently, the main challenge in realizing hybrid task-based quantizers for \ac{mimo} receivers stems from the need to introduce additional processing in analog prior to quantization. Furthermore, this analog combining is required to be dynamically configurable, allowing it to be adapted when operating in dynamic environments. In the following we elaborate on two strategies for implementing such hybrid \ac{mimo} receivers: First, in Section \ref{subsec:PracticeHardware} we discuss hybrid receivers with dedicated analog combining hardware. Then, we present how the emerging technology of \acp{dma} can be exploited to introduce controllable analog combining in Section \ref{subsec:PracticeDMA}.

\vspace{-0.2cm}
\subsection{Dedicated Analog Combiner Hardware}
\label{subsec:PracticeHardware}
\vspace{-0.2cm}
A common strategy to implement \ac{mimo} receivers, particularly when equipped with a large number of antennas and when operating in high spectral bands, is to introduce dedicated analog  circuitry between the antennas and the \acp{adc}. The original motivation for implementing such hybrid receivers is to reduce the number of costly RF chains, namely, the main purpose of the analog combiner is to reduce the dimensionality of the acquired signals allowing the receiver to operate with less RF chains than antennas \cite{mendez2016hybrid,ioushua2019family}. The typical implementation of such analog combiners is based on an inter-connection of phase shifters and adders, either connecting a controllable phase-shifted version of the signal observed at each antenna to each \ac{adc}, resulting in a {\em fully-connected phase shifter network}, or alternatively, by dividing the antennas into subsets, each phase shifted and connected to a distinct \ac{adc}, a topology referred to as {\em partially-connected phase shifter network} \cite{ioushua2019family}. 

The resulting  system model of a hybrid \ac{mimo} receiver thus includes an additional linear processing prior to acquisition, similarly to the model used in our derivation in Section \ref{subsec:TaskLinear}, and can thus be exploited for realizing task-based quantization. In particular, for a hybrid receiver with a fully-connected phase shifter network, the resulting matrix $\myA$ in Section \ref{subsec:TaskLinear} is subject to an additional constraint  which stems from the usage of adjustable phase shifters, that only the phase of its entries can be configured, i.e., $|(\myA)_{i,j}|=1$ for each $i \in \{1,\ldots, \lenZ\}$ and $j \in \{1,\ldots, \lenX\}$. This constraint can accounted for by  identifying the unconstrained analog combining matrix via, e.g., Theorem \ref{thm:OptimalDes},  and projecting it to the feasible set of fully-connected phase shifter networks, similarly to \cite[Alg. 2]{ioushua2019family}. Alternatively, when using a data-driven design as proposed in Section \ref{sec:TaskDeep}, one can account for the additional design constraints by letting the trainable parameters of the analog network to be the phases of the entries of the matrix $\myA$.

\begin{figure}
	\centering
	\includegraphics[width=14cm]{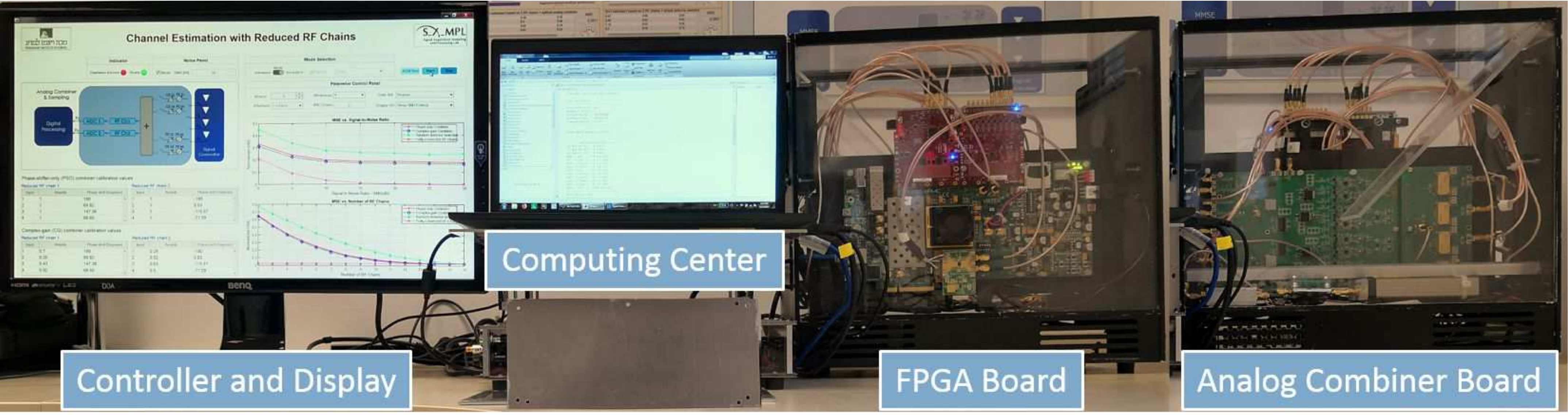} 
	\vspace{-0.2cm}
	\caption{Analog combiner prototype demonstration setup.} 
	\label{fig:PrototypeOverall}
\end{figure}

The difficulties associated with using phase shifter networks as analog combiners for task-based quantization can be mitigated by introducing adjustable gains into the analog circuitry. For example, the prototype proposed in \cite{gong2019rf}, depicted in Fig. \ref{fig:PrototypeOverall}, implements a complex-gain analog combiner operating in the sub-$6$ GHz band using digitally controllable vector multipliers. A controllable gain analog combiner operting in the $25-30$ GHz band based on RF integrated circuits was proposed in \cite{mondal201825}. The resulting model of a hybrid receiver equipped with such analog combiners effectively allows to control both the gain and phase of each entry of the matrix $\myA$ individually in run-time, thus allowing to implement the task-based quantization systems proposed in the previous sections.  The main drawback of such implementations compared to  phase shifter networks is the  cost and complexity associated with controllable complex gain analog circuits.

\vspace{-0.2cm}
\subsection{Analog Combining via Dynamic Metasurface Antennas}
\label{subsec:PracticeDMA}
\vspace{-0.2cm}
The analog combiners discussed in the previous section require the \ac{mimo} receiver to be equipped with a dedicated analog combining hardware interfacing its antenna elements and the \acp{adc}. An alternative strategy to realize configurable analog combining without requiring additional dedicated circuitry is to implement the pre-quantization processing as part of the antenna architecture, by using \acp{dma}. 
The conventional gains of such metasurfaces over architectures based on standard antenna arrays  stem from the fact metasurfaces typically use much less power and cost less  \cite{Johnson-2016TAP}, while facilitating the implementation of a large number of elements in a given physical area. An additional gain of \acp{dma} noted in \cite{shlezinger2019dynamic} is their ability to implement tunable  combining as an inherent byproduct of the antenna architecture. 

\begin{figure}
	\centering
	\includegraphics[width=14cm]{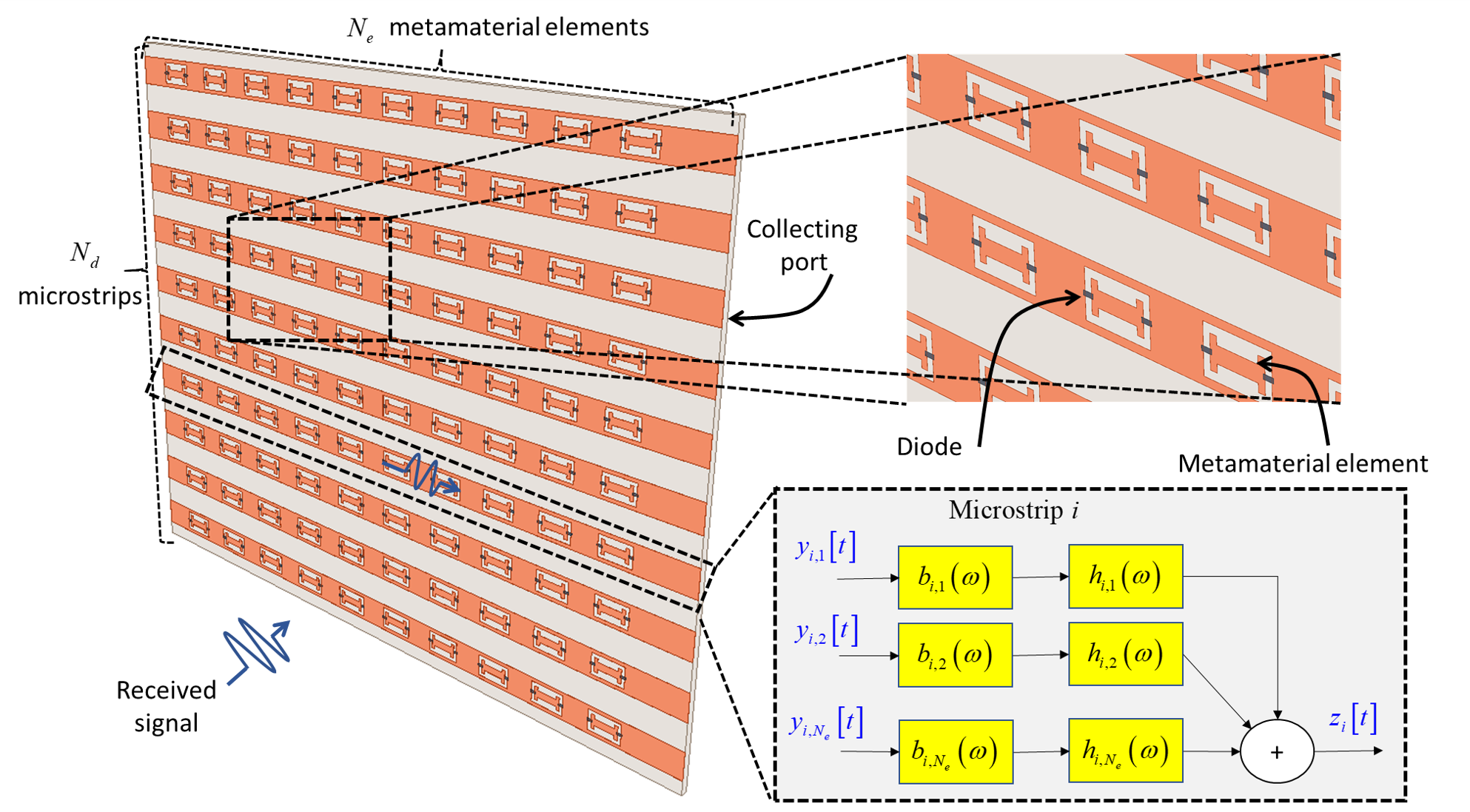} 
	\caption{\ac{dma} system model illustration.} 
	\label{fig:DMA}
\end{figure}

In particular, \acp{dma} consist of a set of microstrips, each embedded with configurable radiating metamaterial elements \cite{DSmith-2017PRA}. When used as a receive antenna, the signals observed by the elements are captured at a single output port for each microstrip, feeding an \ac{adc}. 	The relationship between these signals and the micropstip output is dictated by two main properties: $1)$ Each element of index $l$ of microstrip $i$ acts as resonant electrical circuit, whose frequency response is described by the Lorentzian form \cite{DSmith-2017PRA}
\vspace{-0.2cm}
\begin{equation}\label{eqn:Form_of_Weight}
b_{i,l}(\omega)=\frac{F_{i,l} \omega^2}{(\omega_{i,l}^{R})^{2}-\omega^{2}-j \omega \chi_{i,l}},
\vspace{-0.2cm}
\end{equation}	
where $F_{i,l}$, $\chi_{i,l}$, and $\omega_{i,l}^{R}$ are the  oscillator strength, damping factor, and angular resonance frequency, respectively, which are all externally configurable parameters. 
$2)$ Each signal which propagates from an element to the output port undergoes a different path, and thus accumulates a different delay. The delay accumulated by the signal captured at the $l$th element of the $i$th micropstrip can be modeled as a filter with frequency response $h_{i,l}(\omega)$. The signal observed at the output port of the $i$th micrtopstrip can thus be written as the sum of outputs of the filters $b_{i,l}(\omega)h_{i,l}(\omega)$ whose inputs are the signals observed by the corresponding elements, as illustrated in Fig. \ref{fig:DMA}. 

The resulting model relating the observed signals and the \ac{dma} output ports, which are the signals fed to the \acp{adc}, represents a form of {\em frequency-selective analog combining}. Specifically, the fact that the parameters of the Lorentzian response in \eqref{eqn:Form_of_Weight} can be modified element-wise, indicates that the inherent processing carried out inside each micropstrip can be tuned to facilitate acquisition under bit constraints by tuning the resulting combining as part of a task-based quantizer, see, e.g., \cite{wang2019dynamic}. Consequently, when using a \ac{mimo} receiver with a \ac{dma}-based antenna array, one can implement a form of task-based quantization without requiring additional dedicated analog combining hardware by properly tuning the frequency response of each element along with the quantization mapping and the digital processing utilizing either of the methods discussed in Sections \ref{sec:Task}-\ref{sec:TaskDeep}.

The architectures detailed in this section can all be used to realize task-based quantization in \ac{mimo} receivers, by exploiting either the model-aware design guidelines proposed in Section \ref{sec:Task}, or alternatively, by learning the task-based quantization mapping from labeled data as suggested in Section \ref{sec:TaskDeep}. Combining the architectures detailed in this section with the design methods proposed in the previous sections thus narrows the gap between the theory of task-based quantization and its concrete implementation in \ac{mimo} receivers. 

\vspace{-0.2cm}
\section{Conclusion}
\label{sec:Conclusion}
\vspace{-0.2cm}
In this paper we reviewed the theory and design methods for task-based quantization systems. Such systems carry out acquisition using simple bit-limited scalar \acp{adc}. The associated distortion is mitigated by accounting for the system task in acquisition, via jointly optimizing some level of analog pre-processing along with the quantization rule and digital post-processing in light of the system task. We first presented model-aware design methods which infer the operation of the system components based on prior knowledge of the statistical model relating the observations and the information of interest to be recovered in digital. We then proposed an alternative design approach which does not require knowledge of the underlying model, and learns its task-based quantization mapping from a set of labeled samples using \ac{ml} tools. Finally, we presented several hardware architectures which can facilitate the implementation of task-based quantization mechanisms in \ac{mimo} receivers. The combined results detailed in this survey pave the way to the realization of \ac{mimo} receivers operating accurately and efficiently under strict bit constraints by using task-based quantization techniques.  

\bibliographystyle{IEEEtran}
\bibliography{IEEEabrv,refs}

\end{document}